\documentclass[aps,prc,twocolumn,10pt, superscriptaddress, showpacs, floatfix]{revtex4-1}

\usepackage{amssymb,epsfig}
\usepackage{bbold}

\newcommand{\be}{\begin{equation}}
\newcommand{\ee}{\end{equation}}
\newcommand{\bea}{\begin{eqnarray}}
\newcommand{\eea}{\end{eqnarray}}

\begin{document}

\title{Microscopic response theory for strongly-coupled superfluid fermionic systems}

\author{Elena Litvinova} 
\affiliation{Department of Physics, Western Michigan University, Kalamazoo, MI 49008, USA}
\affiliation{National Superconducting Cyclotron Laboratory, Michigan State University, East Lansing, MI 48824, USA}
\affiliation{GANIL, CEA/DRF-CNRS/IN2P3, F-14076 Caen, France}
\author{Yinu Zhang}
\affiliation{Sino-French Institute of Nuclear Engineering and Technology, Sun Yat-Sen University, Zhuhai, 519082, Guangdong, China}
\affiliation{Department of Physics, Western Michigan University, Kalamazoo, MI 49008, USA}
\date{\today}

\begin{abstract}
A consistent microscopic theory for the response of strongly-coupled superfluid fermionic systems is formulated. 
After defining the response as a two-point two-fermion correlation function in the basis of the Bogolyubov's quasiparticles, the equation of motion (EOM) method is applied  
using the most general fermionic Hamiltonian with a bare two-body interaction, also transformed to the quasiparticle space.
As a superfluid extension of the case of the normal phase, the resulting EOM is of the Bethe-Salpeter-Dyson form with the static and dynamical interaction kernels, where the former determines the short-range correlations and the latter is responsible for the long-range ones.
Both kernels as well as the entire EOM have the double dimension as compared to that of the normal phase. Non-perturbative approximations via the cluster decomposition of the dynamical kernel are discussed, with the major focus on a continuous derivation of the quasiparticle-phonon coupling variant of the latter kernel, where the phonons (vibrations) are composite correlated two-quasiparticle states unifying both the normal and pairing modes.
The developed theory is adopted for nuclear structure applications, such as the nuclear response in various channels. In particular, the finite-amplitude method generalized beyond the quasiparticle random phase approximation, taking into account the quasiparticle-vibration coupling, is formulated for prospective 
calculations in non-spherical nuclei. The article is dedicated to the memory of Peter Schuck and his groundbreaking contributions to the quantum many-body problem.
 
\end{abstract}
%\pacs{21.10.-k, 21.30.Fe, 21.60.-n, 23.40.-s, 24.10.Cn, 24.30.Cz}

\maketitle

%===============================================================================
% Introduction
%===============================================================================
\section{Introduction} 

% Not changed yet

The response function of a finite many-body quantum system is one of its major characteristics, which links, in particular, its interaction 
with the exterior and internal properties. Being defined on the fundamental level as a correlation function of the fermionic field operators, 
it forms the common underlying background connected across the many areas of physics from quantum chromodynamics to quantum chemistry. 
Therefore, understanding microscopic mechanisms of the response, in particular, the phenomenon of emergent collectivity, plays an important role in solving the quantum many-body problem at different scales.

The special role of atomic nuclei in this context stems from the very nature of the nucleon-nucleon (NN) forces. Being the net result of the quark-gluon interaction, it has a complex multichannel structure, which is further tangled by the in-medium dynamics. However, the same feature stipulates the relevance of a large variety of experimental probes, which are sensitive to certain channels and, thus, can illuminate particular aspects of the strong interaction.
These opportunities are actively explored by the rare isotope beam facilities \cite{Tanihata1998,Glasmacher2017}.  

Theoretical studies of the nuclear response were for decades heading toward building a microscopic approach, which can describe nuclear excited states as accurately as possible. For a while, such studies were dominated by the random phase approximation (RPA) \cite{Bohm1951}, or its superfluid variant, the quasiparticle RPA (QRPA) \cite{Baranger1960,RingSchuck1980}, which 
characterize quite adequately the positions and integral strengths of collective excitations, however, produce a poor description of their fragmentation and of other excited states. The response function should be able, in principle, to provide the full spectral picture, therefore, it was realized quite early that the (Q)RPA has to be extended by configurations beyond the two-quasiparticle ($2q$), or one-particle one-hole ($1p1h$), ones. 

The most fruitful non-perturbative extensions of the (Q)RPA appeared with the idea of coupling between the single-particle and emergent collective degrees of freedom in atomic nuclei \cite{BohrMottelson1969,BohrMottelson1975,Broglia1976,BortignonBrogliaBesEtAl1977,BertschBortignonBroglia1983,Soloviev1992}. The nuclear collective modes, phonons, mostly of the vibrational character, were found to form complex configurations by coupling to the single-particle states and to each other. Such configurations cause fragmentation of the (Q)RPA states, thus inducing the damping of collective excitations. 

This idea explained successfully many of the observed nuclear phenomena, while the complex configurations were linked to the dynamical kernels of the equations of motion for the correlation functions in nuclear medium \cite{RingSchuck1980}.
The equation of motion method proposed in Ref. \cite{Rowe1968} and further developed, e.g., in Refs. \cite{Schuck1976,AdachiSchuck1989,Danielewicz1994,DukelskyRoepkeSchuck1998,SchuckTohyama2016,LitvinovaSchuck2019}, is known to produce a hierarchy of approximations to the dynamical kernels of the equations for 
%one-fermion and two-time two-fermion 
fermionic
propagators. Formally, some variants of those kernels for the two-fermion propagators, which are most relevant to the nuclear response, can be mapped to the kernels of the phenomenological nuclear field theories (NFT) and quasiparticle-phonon models (QPM) \cite{BohrMottelson1969,BohrMottelson1975,Broglia1976,BortignonBrogliaBesEtAl1977,BertschBortignonBroglia1983,Soloviev1992}, in terms of $ph\otimes phonon$, $phonon\otimes phonon$ and similar configurations. Such a mapping provides an interpretation of collectivity emerging from the underlying NN-interaction, while in the effective theories of the NFT and QPM types the presence of collective modes and their interactions serves as an input. 

Overall, the EOM method based on the bare Hamiltonian describes the modification of the bare interaction in the strongly-correlated medium. It reveals, in particular, that the latter is not reducible to static 'potentials', but splits, instead, into the static and dynamical components. Both are, in principle, calculable from the underlying interactions, although their ab-initio computations involve evaluations of increasing-rank fermionic propagators: the former is responsible for short-range correlations, while the latter takes into account retardation effects of the correlated media and, thus, long-range correlations \cite{AdachiSchuck1989,Danielewicz1994,DukelskyRoepkeSchuck1998,SchuckTohyama2016,LitvinovaSchuck2019}.

The explicit coupling between the quasiparticles and phonons, commonly referred to as the particle-vibration coupling (PVC) is the leading mechanism of the long-range correlations in nuclear medium. 
Semi-phenomenological models of the PVC based on effective in-medium interactions  \cite{Bortignon1978,Bortignon1981a,BertschBortignonBroglia1983,MahauxBortignonBrogliaEtAl1985,Bortignon1986,Bortignon1997,ColoBortignon2001,Tselyaev1989,KamerdzhievTertychnyiTselyaev1997,Ponomarev2001,Ponomarev1999b,LoIudice2012} provided invaluable knowledge about  this phenomenon in nuclear structure. During the last decades, these models were linked to the contemporary density functional theories in self-consistent frameworks \cite{LitvinovaRing2006,LitvinovaRingTselyaev2008,LitvinovaRingTselyaev2010,Tselyaev2013,AfanasjevLitvinova2015,Tselyaev2016,NiuNiuColoEtAl2015,Niu2016,Niu2018} and widely applied to experimental data analyses \cite{EndresLitvinovaSavranEtAl2010,PoltoratskaFearickKrumbholzEtAl2014,MassarczykSchwengnerDoenauEtAl2012,LanzaVitturiLitvinovaEtAl2014,Oezel-TashenovEndersLenskeEtAl2014}. Furthermore, the PVC with charge-exchange phonons \cite{Litvinova2016,RobinLitvinova2018} and the PVC-induced ground state correlations \cite{Kamerdzhiev1991,Tselyaev1989,KamerdzhievTertychnyiTselyaev1997,Robin2019}, have been addressed in applications to neutral and charge-exchange nuclear excitations.  Some recent developments and numerical implementations of the NFT's beyond the two-particle two-hole $2p2h$ level \cite{Tselyaev2018,Litvinova2015,LitvinovaSchuck2019}, as well as the multiphonon QPM approach \cite{Soloviev1992,Ponomarev2001,SavranBabilonBergEtAl2006,Andreozzi2008} indicate the capability of a sufficiently advanced response theory to meet the shell-model standards in large model spaces. 

However, the problem of consistent linking those approaches to the nuclear response with the underlying bare interactions remains unsolved, although some theoretical effort in this direction was made recently in Refs. \cite{LitvinovaSchuck2019,LitvinovaSchuck2020,Schuck2021,Litvinova2021}. 
Another aspect, which needs a special attention, is the superfluidity, which famously manifests itself in open-shell nuclei. Superfluid dynamical PVC kernels of the EOM for the response function were considered in a number of reports \cite{Soloviev1992,Ponomarev2001,ColoBortignon2001,Tselyaev2007,LitvinovaTselyaev2007,LitvinovaRingTselyaev2008,RobinLitvinova2016,Niu2016}, however, pairing correlations were only taken into account in the Bardeen-Cooper-Schrieffer (BCS) approximation. Meanwhile, it became evident that a more accurate treatment of superfluidity is highly desirable for modern nuclear physics applications. 

In this work, an advancement of the response theory with the dynamical kernel is presented for superfluid fermionic systems.
The pairing correlations are treated consistently by working in the Hartree-Fock-Bogolyubov (HFB) basis of the Bogolyubov's quasiparticles, i.e., beyond the BCS pairing. Starting from a general many-body fermionic Hamiltonian defined by the bare two-body interaction, the EOM formalism for the response function is processed in the HFB space. A connection to the variational approach is made to guide future applications of the method to a broad class of atomic nuclei. Namely, we show how the finite-amplitude method (FAM), which is extremely efficient for deformed nuclei but, up until now formulated and implemented only on the QRPA level \cite{Avogadro2011,Avogadro2013,Hinohara2013,Niksic2013,Kortelainen2015,Bjelcic2020}, can be generalized to include correlations beyond the QRPA. These correlations are responsible for many important nuclear characteristics, such as the neutron capture and weak rates in stars, neutrino-nucleus interaction, neutrinoless double beta decay, nuclear Schiff moment, i.e., the quantities needed for major frontier applications. These quantities are extremely sensitive to the fine structure of the nuclear excitation spectra, which can not be described by
the QRPA, in principle. The presented approach and, in particular, its FAM variant generalized beyond the QRPA will, besides advancing the theory alone, allow for highly-accurate computation of these quantities in a wide range of nuclear masses, isospins and shapes. 

The main purpose of this article is to walk the reader throughout the detailed formalism and, thus, provide a clear understanding of the origin of the many-body effects, which contribute to the response of a superfluid fermionic system. Since the static kernel of the EOM for the superfluid response function is, in the leading approximation, known from the QRPA \cite{RingSchuck1980}, we place the major focus on the dynamical kernel, which is responsible for more non-trivial, relatively long-range, correlations emerging from the strongly-coupled superfluid medium in finite volume. While we analyze explicitly two specific approaches to the dynamical kernel, the model-independent nature of the exact EOM should allow practitioners to relate different many-body models to each other, to evaluate their accuracy and generate systematically improvable approximations. We purposefully do not include numerical calculations here to fully focus on the formalism, as it is done, for instance, in Refs. \cite{Tselyaev2007,Soma2011,Litvinova2021}, some aspects of which can be related to the presented theory in certain approximations, while a separate series of articles will be reserved for its numerical implementations.

%===============================================================================
% Formalism
%===============================================================================
\section{Formalism}
\label{Propagators}
\subsection{Many-body Hamiltonian in the quasiparticle basis}

The starting point for the fermionic EOM formalism is the Hamiltonian $H$ of a many-body system of interacting fermions. It is conventionally formulated as
\be
H = H^{(1)} + V^{(2)},
\label{Hamiltonian}
\ee
where $H^{(1)}$  and $V^{(2)}$ are the one-body and two-body parts, respectively. Higher-rank terms are neglected in the present work as their contribution is usually less important, however, they can be included straightforwardly, if necessary. 
The operator $H^{(1)}$, in terms of the fermionic field operators, reads
\be
H^{(1)} = \sum_{12} t_{12} \psi^{\dag}_1\psi_2 + \sum_{12}v^{(MF)}_{12}\psi^{\dag}_1\psi_2 \equiv \sum_{12}h_{12}\psi^{\dag}_1\psi_2,
\label{Hamiltonian1}
\ee
where the matrix elements $h_{12}$ combine the kinetic energy $t$ and the mean-field $v^{(MF)}$ portion of the interaction. The external mean field, in case of its presence, is included in $v^{(MF)}$. The two-fermion interaction $V^{(2)}$ is 
\be
V^{(2)} = \frac{1}{4}\sum\limits_{1234}{\bar v}_{1234}\psi^{\dagger}_1\psi^{\dagger}_2\psi_4\psi_3,
\label{Hamiltonian2}
\ee
where ${\bar v}_{1234} = v_{1234} - v_{1243}$ are the antisymmetrized matrix elements of the interaction between two fermions in free space. We do not specify the nature of the interaction $v$ assuming only its instantaneous character. This assumption should be quite accurate in the low-energy regime for the majority of the fermionic systems of interest. Non-instantaneous interactions as well as the three-body and higher-rank forces will be discussed elsewhere. 
The number lower indices in Eqs. (\ref{Hamiltonian1}) and (\ref{Hamiltonian2}) stand for the complete sets of quantum numbers in an arbitrary basis, and the fermionic field operators $\psi_1$ and $\psi^{\dagger}_1$ satisfy the usual anticommutation relations. 

% Put a note about convenience of the quasiparticle basis, while in the canonical basis there would be 16 propagators to calculate.
For the description of a superfluid system of fermions, it is convenient to proceed in the space of Bogolyubov's quasiparticles defined by the transformation \cite{Bogolyubov1958,RingSchuck1980}
\bea
\psi_1 = U_{1\mu}\alpha_{\mu} + V^{\ast}_{1\mu}\alpha^{\dagger}_{\mu}\nonumber\\
\psi^{\dagger}_1 = V_{1\mu}\alpha_{\mu} + U^{\ast}_{1\mu}\alpha^{\dagger}_{\mu},
\label{Btrans}
\eea
where summation is implied over the repeated index $\mu$, or, in the operator form:
\bea
\left( \begin{array}{c} \psi \\ \psi^{\dagger} \end{array} \right) = \cal{W} \left( \begin{array}{c} \alpha \\ \alpha^{\dagger} \end{array} \right),
\eea
where
\bea
\cal{W} = \left( \begin{array}{cc} U & V^{\ast} \\ V & U^{\ast} \end{array} \right) \ \ \ \ \ \  \cal{W}^{\dagger} = \left( \begin{array}{cc} U^{\dagger} & V^{\dagger} \\ V^T & U^T \end{array} \right). 
\eea
In Eq. (\ref{Btrans}) and henceforth the Greek subscripts will be used to denote fermionic states in the HFB basis, while the number subscripts 
%and the Roman ones introduced below 
will remain reserved for the single-particle (mean-field) basis states.
The transformation $\cal W$ is unitary, and the quasiparticle operators $\alpha$ and $\alpha^{\dagger}$ form the same anticommutator algebra as the particle operators $\psi$ and $\psi^{\dagger}$, so that the matrices $U$ and $V$ satisfy:
\bea
U^{\dagger}U + V^{\dagger}V = \mathbb{1}\ \ \ \ \ \ UU^{\dagger} + V^{\ast}V^{T} = \mathbb{1}\nonumber\\
U^TV + V^TU = 0\ \ \ \ \ \  UV^{\dagger} + V^{\ast}U^{T} = 0 .
\label{UV}
\eea
  
In the quasiparticle basis defined by Eq. (\ref{Btrans}), the Hamiltonian (\ref{Hamiltonian}) can be rewritten as follows \cite{RingSchuck1980}:
\bea
H = H^0 &+& \sum\limits_{\mu\nu}H^{11}_{\mu\nu}\alpha^{\dagger}_{\mu}\alpha_{\nu}  + \frac{1}{2}\sum\limits_{\mu\nu}\bigl(H^{20}_{\mu\nu}\alpha^{\dagger}_{\mu}\alpha^{\dagger}_{\nu} + \text{h.c.}\bigr) \nonumber\\
&+& \sum\limits_{\mu\mu'\nu\nu'}\bigl(H^{40}_{\mu\mu'\nu\nu'}\alpha^{\dagger}_{\mu}\alpha^{\dagger}_{\mu'}\alpha^{\dagger}_{\nu}\alpha^{\dagger}_{\nu'} + \text{h.c}\bigr) \nonumber\\
&+& \sum\limits_{\mu\mu'\nu\nu'}\bigl(H^{31}_{\mu\mu'\nu\nu'}\alpha^{\dagger}_{\mu}\alpha^{\dagger}_{\mu'}\alpha^{\dagger}_{\nu}\alpha_{\nu'} + \text{h.c}\bigr) 
\nonumber\\
&+& \frac{1}{4}\sum\limits_{\mu\mu'\nu\nu'}\bigl(H^{22}_{\mu\mu'\nu\nu'}\alpha^{\dagger}_{\mu}\alpha^{\dagger}_{\mu'}\alpha_{\nu'}\alpha_{\nu} + \text{h.c}\bigr), 
\label{Hqua}
\eea
where the upper indices in the matrix elements $H^{ij}_{\mu\nu\mu'\nu'}$ are associated with the numbers of creation and annihilation quasiparticle operators in the respective terms. These matrix elements are given in Appendix \ref{AppA}. 

While the matrix $H^{20}$ vanishes at the stationary point defining the Hartree-Fock-Bogolyubov equations, the matrix elements of $H^{11}$ correspond to the quasiparticle energies, so that $H^{11}_{\mu\nu} = \delta_{\mu\nu}E_{\mu}$, and the Hamiltonian takes the form
 \cite{RingSchuck1980}:
 \be
 H = H^0 + \sum\limits_{\mu}E_{\mu}\alpha^{\dagger}_{\mu}\alpha_{\mu} + V,
 \label{Hqua1}
 \ee
where the residual interaction $V$ includes the $H^{40}$, $H^{31}$ and $H^{22}$ terms of Eq. (\ref{Hqua}):
 \bea
 V &=& 
 \sum\limits_{\mu\mu'\nu\nu'}\bigl(H^{40}_{\mu\mu'\nu\nu'}\alpha^{\dagger}_{\mu}\alpha^{\dagger}_{\mu'}\alpha^{\dagger}_{\nu}\alpha^{\dagger}_{\nu'} + \text{h.c}\bigr) \nonumber\\
&+& \sum\limits_{\mu\mu'\nu\nu'}\bigl(H^{31}_{\mu\mu'\nu\nu'}\alpha^{\dagger}_{\mu}\alpha^{\dagger}_{\mu'}\alpha^{\dagger}_{\nu}\alpha_{\nu'} + \text{h.c}\bigr) 
\nonumber\\
&+& \frac{1}{4}\sum\limits_{\mu\mu'\nu\nu'}\bigl(H^{22}_{\mu\mu'\nu\nu'}\alpha^{\dagger}_{\mu}\alpha^{\dagger}_{\mu'}\alpha_{\nu'}\alpha_{\nu} + \text{h.c}\bigr).
\label{Hqua2} 
\eea

\subsection{Strength function and superfluid response}

The response of a fermionic system to an external field $F$ can be deduced from the associated generic strength function which, under the assumption of a sufficiently weak field reads:
\be
S(\omega) = \sum\limits_{n>0} \Bigl[ |\langle n|F^{\dagger}|0\rangle |^2\delta(\omega-\omega_n) - |\langle n|F|0\rangle |^2\delta(\omega+\omega_n)
\Bigr],
\label{SF}
\ee
where the summation runs over all the formally exact excited states $|n\rangle$. The square moduli of the matrix elements express the transition probabilities \footnote{Note that in Ref. \cite{Litvinova2021} the operators $F$ and $F^{\dagger}$ were interchanged, that is a matter of convention.}
\be
B_n = |\langle n|F^{\dagger}|0\rangle |^2\ \ \ \ \ \ \ \ 
{\bar B}_n = |\langle n|F|0\rangle |^2.
\label{Prob}
\ee

Considering a superfluid system, it is convenient to represent the generic one-body operator $F$ in terms of the quasiparticle fields:
\bea
F = \frac{1}{2}\sum\limits_{\mu\mu'} \Bigl(F^{20}_{\mu\mu'}\alpha^{\dagger}_{\mu}\alpha^{\dagger}_{\mu'} + 
F^{02}_{\mu\mu'}\alpha_{\mu'}\alpha_{\mu} \Bigr)\nonumber\\
F^{\dagger} = \frac{1}{2}\sum\limits_{\mu\mu'} \Bigl(F^{20\ast}_{\mu\mu'}\alpha_{\mu'}\alpha_{\mu} +
F^{02\ast}_{\mu\mu'}\alpha^{\dagger}_{\mu}\alpha^{\dagger}_{\mu'}  
\Bigr),
\label{Fext}
\eea
that follows from the Bogolyubov's transformation of the second-quantized form of $F$. The full composition in the quasiparticle basis contains formally also $F^{11}$ terms, however, their contribution vanishes in the leading approximations to the superfluid response, for instance, in the zero-temperature QRPA \cite{Avogadro2011}. The associated contributions to the response function are related to complex ground state correlations and will be considered elsewhere.
Here we also note that $F^{02\ast}_{\mu\mu'} = F^{20}_{\mu\mu'}$, if $F = F^{\dagger}$, i.e., a Hermitian operator.

While Eq. (\ref{SF}) is model independent, the matrix elements in it obviously depend on the model assumptions about both the ground $|0\rangle$ and excited $|n\rangle$ states. The generic structure of the transition probabilities can be recast as
\bea
|\langle n|F^{\dagger}|0\rangle |^2 = \frac{1}{4}\sum\limits_{\mu\mu'\nu\nu'}
\left(\begin{array}{cc} F^{02}_{\mu\mu'} & F^{20}_{\mu\mu'} \end{array}\right) \nonumber\\
\times\left(\begin{array}{cc} {\cal X}^{n}_{\mu\mu'}{\cal X}^{n\ast}_{\nu\nu'}  &  {\cal X}^{n}_{\mu\mu'}{\cal Y}^{n\ast}_{\nu\nu'}
\\
{\cal Y}^{n}_{\mu\mu'}{\cal X}^{n\ast}_{\nu\nu'}  &  {\cal Y}^{n}_{\mu\mu'}{\cal Y}^{n\ast}_{\nu\nu'}\end{array}\right)
\left(\begin{array}{c} F^{02\ast}_{\nu\nu'} \\ F^{20\ast}_{\nu\nu'} \end{array}\right)
\label{nFd0}
\eea
and
\bea
|\langle n|F|0\rangle |^2 = \frac{1}{4}\sum\limits_{\mu\mu'\nu\nu'}
\left(\begin{array}{cc} F^{02\ast}_{\mu\mu'} & F^{20\ast}_{\mu\mu'} \end{array}\right) \nonumber\\
\times\left(\begin{array}{cc} {\cal Y}^{n}_{\mu\mu'}{\cal Y}^{n\ast}_{\nu\nu'}  &  {\cal Y}^{n}_{\mu\mu'}{\cal X}^{n\ast}_{\nu\nu'}
\\
{\cal X}^{n}_{\mu\mu'}{\cal Y}^{n\ast}_{\nu\nu'}  &  {\cal X}^{n}_{\mu\mu'}{\cal X}^{n\ast}_{\nu\nu'}\end{array}\right)
\left(\begin{array}{c} F^{02}_{\nu\nu'} \\ F^{20}_{\nu\nu'} \end{array}\right),
\label{nF0}
\eea
via the matrix elements
\be
{\cal X}^{n}_{\mu\mu'} = \langle 0|\alpha_{\mu'}\alpha_{\mu}|n\rangle \ \ \ \ \ \ \ \ \ \ {\cal Y}^{n}_{\mu\mu'} = \langle 0|\alpha^{\dagger}_{\mu}\alpha^{\dagger}_{\mu'}|n\rangle .
\label{XY}
\ee
Note here that all the matrix elements in the right hand sides of Eqs. (\ref{Fext} - \ref{XY}) are antisymmetric because of the fermionic character of the quasiparticles. Therefore, each factor $1/2$ in front of the summation over a couple of the fermionic states $\{\mu,\mu'\}$ can be removed by restricting the sum by $\mu < \mu'$. Since for the antisymmetric matrices the diagonal matrix elements vanish, the condition $\mu < \mu'$ is equivalent to $\mu \leq \mu'$. The non-vanishing "diagonal" matrix elements may appear in certain implementations, where the single-quasiparticle states are degenerate with respect to one or several quantum numbers of the complete set $\{\mu\}$ and if the dependence on these quantum numbers is explicitly excluded \cite{LitvinovaTselyaev2007,LitvinovaRingTselyaev2008,Niu2016}.
In the following, depending on the context, we will use the complete or restricted summations, which can be easily converted to each other, if necessary.  

The $\delta$-function in Eq. (\ref{SF}) can be represented as a limit of the finite-width Lorentzian distribution
\be
\delta(\omega-\omega_n) = \frac{1}{\pi}\lim\limits_{\Delta \to 0}\frac{\Delta}{(\omega - \omega_n)^2 + \Delta^2},
\ee
so that
\bea
S(\omega) = \frac{1}{\pi}\lim\limits_{\Delta \to 0}\sum\limits_{n>0} \Bigl[ |\langle n|F^{\dagger}|0\rangle |^2\frac{\Delta}{(\omega - \omega_n)^2 + \Delta^2}
- \nonumber \\
- |\langle n|F|0\rangle |^2\frac{\Delta}{(\omega + \omega_n)^2 + \Delta^2}
\Bigr] = \nonumber\\
= -\frac{1}{\pi}\lim\limits_{\Delta \to 0} \text{Im}\sum\limits_{n>0} \Bigl[\frac{ |\langle n|F^{\dagger}|0\rangle |^2}{\omega - \omega_n + i\Delta}
- \frac{|\langle n|F|0\rangle |^2}{\omega + \omega_n + i\Delta}
\Bigr]= \nonumber\\
= -\frac{1}{\pi}\lim\limits_{\Delta \to 0} \text{Im} \Pi(\omega).\nonumber\\
\label{SFDelta} 
\eea
The polarizability $\Pi(\omega)$ of the system is, thus, defined as
\bea
\Pi(\omega) = \sum\limits_{n>0} \Bigl[ \frac{|\langle n|F^{\dagger}|0\rangle |^2}{\omega - \omega_n + i\Delta}
- \frac{|\langle n|F|0\rangle |^2}{\omega + \omega_n + i\Delta}
\Bigr] \nonumber\\
 = \frac{1}{4}\sum\limits_{\mu\mu'\nu\nu'}
\left(\begin{array}{cc} F^{02}_{\mu\mu'} & F^{20}_{\mu\mu'} \end{array}\right) \nonumber\\
\times\left(\begin{array}{cc} {R}^{[11]}_{\mu\mu'\nu\nu'}(\omega+i\Delta)  &  {R}^{[12]}_{\mu\mu'\nu\nu'}(\omega+i\Delta) 
\\
\\
{R}^{[21]}_{\mu\mu'\nu\nu'}(\omega+i\Delta)  & {R}^{[22]}_{\mu\mu'\nu\nu'}(\omega+i\Delta) \end{array}\right)
\left(\begin{array}{c} F^{02\ast}_{\nu\nu'} \\  \\ F^{20\ast}_{\nu\nu'} \end{array}\right),\nonumber\\
\label{Polar}
\eea
where the matrix elements of the response function ${\hat{\cal R}}_{\mu\mu'\nu\nu'}(\omega) = \{R^{[ij]}_{\mu\mu'\nu\nu'}(\omega)\}$ with $i,j = \{1,2\}$ read:
\bea
{\hat{\cal R}}_{\mu\mu'\nu\nu'}(\omega) = \nonumber \\
= \sum\limits_{n>0} \left(\begin{array}{c} {\cal X}^{n}_{\mu\mu'} \\ {\cal Y}^{n}_{\mu\mu'} \end{array}\right)
\frac{1}{\omega - \omega_n}\left(\begin{array}{cc} {\cal X}^{n\ast}_{\nu\nu'} & {\cal Y}^{n\ast}_{\nu\nu'} \end{array}\right)\nonumber \\
- \sum\limits_{n>0} \left(\begin{array}{c} {\cal Y}^{n\ast}_{\mu\mu'} \\ {\cal X}^{n\ast}_{\mu\mu'} \end{array}\right)
\frac{1}{\omega + \omega_n}\left(\begin{array}{cc} {\cal Y}^{n}_{\nu\nu'} & {\cal X}^{n}_{\nu\nu'} \end{array}\right).\nonumber \\
\label{Romega}
\eea
To apply the EOM formalism, it is convenient to formulate the superfluid response function in terms of the time-dependent field operators, in analogy to the normal case. 
The definition should be compatible with Eq. (\ref{Romega}), which represents the Fourier image of the superfluid response function. Thus, we adopt the 
definition
\bea
&{\hat{\cal R}}&_{\mu\mu'\nu\nu'} (t-t') = -i\nonumber
\\
&\times&\left(\begin{array}{cc}
\langle T(\alpha_{\mu'}\alpha_{\mu})(t)(\alpha^{\dagger}_{\nu}\alpha^{\dagger}_{\nu'})(t')\rangle & 
\langle T(\alpha_{\mu'}\alpha_{\mu})(t)(\alpha_{\nu'}\alpha_{\nu})(t')\rangle \\
\langle T(\alpha^{\dagger}_{\mu}\alpha^{\dagger}_{\mu'})(t)(\alpha^{\dagger}_{\nu}\alpha^{\dagger}_{\nu'})(t')\rangle &
\langle T(\alpha^{\dagger}_{\mu}\alpha^{\dagger}_{\mu'})(t)(\alpha_{\nu'}\alpha_{\nu})(t')\rangle
\end{array}\right) ,\nonumber\\
\label{Rt}
\eea
where the time-dependent operator products should be understood as those in the Heisenberg picture:
\bea
(\alpha_{\mu'}\alpha_{\mu})(t) = e^{iHt}\alpha_{\mu'}\alpha_{\mu}e^{-iHt} \nonumber \\ 
(\alpha^{\dagger}_{\nu}\alpha^{\dagger}_{\nu'})(t) = 
e^{iHt}\alpha^{\dagger}_{\nu}\alpha^{\dagger}_{\nu'}e^{-iHt}
\eea 
with the $\hbar = 1$ convention, $T$ is the time ordering operator, and the averaging is performed over the formally exact ground state. For processing the response function ${\hat{\cal R}}_{\mu\mu'\nu\nu'} (t-t')$ in the EOM framework, it is useful to recast it in terms of the quasiparticle pair operators, namely:
\bea
&{\hat{\cal R}}&_{\mu\mu'\nu\nu'} (t-t') = -i\nonumber \\
&\times&\left(\begin{array}{cc}
\langle TA_{\mu\mu'}(t)A^{\dagger}_{\nu\nu'}(t')\rangle  &
\langle TA_{\mu\mu'}(t)A_{\nu\nu'}(t')
\rangle \\
\langle TA^{\dagger}_{\mu\mu'}(t)A^{\dagger}_{\nu\nu'}(t')\rangle &
\langle TA^{\dagger}_{\mu\mu'}(t)A_{\nu\nu'}(t')\rangle
\end{array}\right),\nonumber\\
\label{RtA}
\eea 
where the latter operators are introduced according to:
\be
A_{\mu\mu'} = \alpha_{\mu'}\alpha_{\mu} \ \ \ \ \ \ \ \ \ \ A^{\dagger}_{\mu\mu'} = \alpha^{\dagger}_{\mu}\alpha^{\dagger}_{\mu'}.
\label{Aq}
\ee

Indeed, inserting the operator unit $I = \sum_n|n\rangle\langle n|$, in terms of the full set of the eigenstates $|n\rangle$ of the Hamiltonian $H$, between the quasiparticle pair operators in each of the matrix elements, one gets, after the Fourier transformation to the frequency (energy) domain:
\bea
{\hat{\cal R}}_{\mu\mu'\nu\nu'}(\omega) = \sum\limits_{n>0}
\left(\begin{array}{cc} {\cal X}^{n}_{\mu\mu'}{\cal X}^{n\ast}_{\nu\nu'}  &  {\cal X}^{n}_{\mu\mu'}{\cal Y}^{n\ast}_{\nu\nu'}
\\
{\cal Y}^{n}_{\mu\mu'}{\cal X}^{n\ast}_{\nu\nu'}  &  {\cal Y}^{n}_{\mu\mu'}{\cal Y}^{n\ast}_{\nu\nu'}\end{array}\right)
\frac{1}{\omega-\omega_n+i\delta} \nonumber\\
-\sum\limits_{n>0}\left(\begin{array}{cc} {\cal Y}^{n\ast}_{\mu\mu'}{\cal Y}^{n}_{\nu\nu'}  &  {\cal Y}^{n\ast}_{\mu\mu'}{\cal X}^{n}_{\nu\nu'}
\\
{\cal X}^{n\ast}_{\mu\mu'}{\cal Y}^{n}_{\nu\nu'}  &  {\cal X}^{n\ast}_{\mu\mu'}{\cal X}^{n}_{\nu\nu'}\end{array}\right)
\frac{1}{\omega+\omega_n-i\delta}, \nonumber\\
\label{R}
\eea
where $\delta\to +0$ is the infinitesimal imaginary part of the energy variable introduced for convergence of the integrals for both the retarded and advanced contributions. Thus, one can see that Eq. (\ref{Romega}) is reproduced with the definition of Eq. (\ref{Rt}), if the different origins of the imaginary parts $\Delta$ and $\delta$ of the energy variable are taken into account. Remarkably, the response function in the form of Eq. (\ref{RtA}) has the same structure as the single-quasiparticle Gorkov propagator discussed within the EOM framework, in particular, in Refs. \cite{Litvinova2021,Zhang2022}, with the correspondence $\psi_1 \to A_{\mu\mu'}$, $\psi^{\dagger}_1 \to A^{\dagger}_{\mu\mu'}$. It should be, thus, straightforward to handle it with an analogous EOM method.

\subsection{Equation of motion for the superfluid response}

The time derivative $\partial_t$ of the superfluid response function (\ref{RtA}) reads:
\bea
\partial_t{\hat{\cal R}}_{\mu\mu'\nu\nu'}(t-t') = -i\delta(t-t') \nonumber\\
\times\left(\begin{array}{cc}\langle[A_{\mu\mu'},A^{\dagger}_{\nu\nu'}]\rangle  & \langle[A_{\mu\mu'},A_{\nu\nu'}]\rangle 
\\
\langle [A^{\dagger}_{\mu\mu'},A^{\dagger}_{\nu\nu'}]\rangle  &
\langle [A^{\dagger}_{\mu\mu'},A_{\nu\nu'}]\rangle
\end{array}\right) \nonumber\\
+ \left(\begin{array}{cc}\langle T[H,A_{\mu\mu'}](t)A^{\dagger}_{\nu\nu'}(t')\rangle  & \langle T[H,A_{\mu\mu'}](t)A_{\nu\nu'}(t')\rangle
\\
\langle T[H,A^{\dagger}_{\mu\mu'}](t)A^{\dagger}_{\nu\nu'}(t')\rangle  &
\langle T[H,A^{\dagger}_{\mu\mu'}](t)A_{\nu\nu'}(t')\rangle
\end{array}\right).\nonumber\\
\label{dR0}
\eea
%so that one can immediately recognize the 
At this step, when one foresees evaluation of the commutators appearing in the EOM, it is convenient to separate the $H^0$ and $H^{11}$ parts of the full Hamiltonian (\ref{Hqua}) from the remaining terms as in Eq. (\ref{Hqua1}). Thus, one finds explicitly that
\bea
[H,A_{\mu\mu'}] = -(E_{\mu} + E_{\mu'})A_{\mu\mu'} + [V,A_{\mu\mu'}]\nonumber\\
\left[H,A^{\dagger}_{\mu\mu'}\right] = (E_{\mu} + E_{\mu'} )A^{\dagger}_{\mu\mu'} + [V,A^{\dagger}_{\mu\mu'}],
\label{Comm1}
\eea
and Eq. (\ref{dR0}) leads to the first EOM, which can be written as follows:
\bea
\left[i\partial_t - \left(\begin{array}{cc} 1& 0 \\ 0 & -1 \end{array}\right)(E_{\mu} + E_{\mu'})\right]{\hat{\cal R}}_{\mu\mu'\nu\nu'}(t-t') 
\ \ \ \ \ \ \ \ \ \ \ \ \
 \nonumber\\
= \delta(t-t'){\hat{\cal N}}_{\mu\mu'\nu\nu'} \ \ \ \ \ \ \ \ \ \ \ \ \ \ \ \ \ \ \ \ \ \ \ \ \ \ \ \ \ \ \ \ \ \ \ \ \ \ \ \ \ \ \ \ \ \  \nonumber\\
+ i\left(\begin{array}{cc}\langle T[V,A_{\mu\mu'}](t)A^{\dagger}_{\nu\nu'}(t')\rangle  & \langle T[V,A_{\mu\mu'}](t)A_{\nu\nu'}(t')\rangle
\\
\langle T[V,A^{\dagger}_{\mu\mu'}](t)A^{\dagger}_{\nu\nu'}(t')\rangle  &
\langle T[V,A^{\dagger}_{\mu\mu'}](t)A_{\nu\nu'}(t')\rangle 
\end{array}\right),\nonumber\\
\label{EOM1}
\eea
where the norm matrix ${\hat{\cal N}}_{\mu\mu'\nu\nu'}$ is defined as:
\be
{\hat{\cal N}}_{\mu\mu'\nu\nu'} = \left(\begin{array}{cc}\langle[A_{\mu\mu'},A^{\dagger}_{\nu\nu'}]\rangle  & 0
\\
0  &
\langle[A^{\dagger}_{\mu\mu'},A_{\nu\nu'}]\rangle
\end{array}\right).
\label{norm}
\ee

Analogously to the non-superfluid case, the second EOM is generated for the dynamical term on the right hand side of Eq. (\ref{EOM1}). Denoting this term as
\bea
{\hat{\cal F}}_{\mu\mu'\nu\nu'}(t-t') =
\nonumber \\ = i\left(\begin{array}{cc}\langle T[V,A_{\mu\mu'}](t)A^{\dagger}_{\nu\nu'}(t')\rangle  & \langle T[V,A_{\mu\mu'}](t)A_{\nu\nu'}(t')\rangle
\\
\langle T[V,A^{\dagger}_{\mu\mu'}](t)A^{\dagger}_{\nu\nu'}(t')\rangle  &
\langle T[V,A^{\dagger}_{\mu\mu'}](t)A_{\nu\nu'}(t')\rangle
\end{array}\right) \nonumber \\
\label{F}
\eea
and taking its derivative with respect to $t'$ leads to
\bea
\partial_{t'}{\hat{\cal F}}_{\mu\mu'\nu\nu'}(t-t') = -i\delta(t-t') \nonumber\\
\times\left(\begin{array}{cc}\langle[[V,A_{\mu\mu'}],A^{\dagger}_{\nu\nu'}]\rangle  & \langle[[V,A_{\mu\mu'}],A_{\nu\nu'}]\rangle 
%\\
\\
\langle[[V,A^{\dagger}_{\mu\mu'}],A^{\dagger}_{\nu\nu'}]\rangle  &
\langle[[V,A^{\dagger}_{\mu\mu'}],A_{\nu\nu'}]\rangle
\end{array}\right) \nonumber\\
- \left(\begin{array}{cc}\langle T[V,A_{\mu\mu'}](t)[H,A^{\dagger}_{\nu\nu'}](t')\rangle  & \langle T[V,A_{\mu\mu'}](t)[H,A_{\nu\nu'}](t')\rangle
\\
\langle T[V,A^{\dagger}_{\mu\mu'}](t)[H,A^{\dagger}_{\nu\nu'}](t')\rangle  &
\langle T[V,A^{\dagger}_{\mu\mu'}](t)[H,A_{\nu\nu'}](t')\rangle
\end{array}\right).\nonumber\\
\label{dR}
\eea
After processing the commutators with the Hamiltonian according to Eq. (\ref{Comm1}), the second EOM takes the form:
\bea
{\hat{\cal F}}_{\mu\mu'\nu\nu'}(t&-&t') \left[-i\overleftarrow{\partial_{t'}} - (E_{\nu} + E_{\nu'})\left(\begin{array}{cc} 1& 0 \\ 0 & -1 \end{array}\right)\right]
=
 \nonumber\\
&=& \delta(t-t'){\hat{\cal T}}^{0}_{\mu\mu'\nu\nu'} + {\hat{\cal T}}^{r}_{\mu\mu'\nu\nu'}(t-t'),
\label{EOM2}
\eea
where ${\hat{\cal T}}^{0}_{\mu\mu'\nu\nu'}$  and ${\hat{\cal T}}^{r}_{\mu\mu'\nu\nu'}(t-t')$ define the static and dynamical (retarded) parts of the 
two-fermion ${\hat{\cal T}}$-matrix in the quasiparticle space:
\be
{\hat{\cal T}}^{0}_{\mu\mu'\nu\nu'} = -\left(\begin{array}{cc}\langle[[V,A_{\mu\mu'}],A^{\dagger}_{\nu\nu'}]\rangle  & 
\langle[[V,A_{\mu\mu'}],A_{\nu\nu'}]\rangle 
%\\
\\
\langle[[V,A^{\dagger}_{\mu\mu'}],A^{\dagger}_{\nu\nu'}]\rangle  &
\langle[[V,A^{\dagger}_{\mu\mu'}],A_{\nu\nu'}]\rangle
\end{array}\right)
\label{T0}
\ee
\bea
{\hat{\cal T}}^{r}_{\mu\mu'\nu\nu'}(t-t') = i\times\ \ \ \ \ \ \ \ \ \ \ \ \ \ \ \ \ \ \ \ \ \ \ \ \ \ \ \ \ \ \ \ \ \ \ \ \ \ \ \ \ \ \ \ \ \
\nonumber\\ \times
\left(\begin{array}{cc}\langle T[V,A_{\mu\mu'}](t)[V,A^{\dagger}_{\nu\nu'}](t')\rangle  & \langle T[V,A_{\mu\mu'}](t)[V,A_{\nu\nu'}](t')\rangle
\\
\langle T[V,A^{\dagger}_{\mu\mu'}](t)[V,A^{\dagger}_{\nu\nu'}](t')\rangle  &
\langle T[V,A^{\dagger}_{\mu\mu'}](t)[V,A_{\nu\nu'}](t')\rangle
\end{array}\right). \nonumber\\
\label{Tr}
\eea
Combining Eqs. (\ref{EOM1}) and (\ref{EOM2}), namely acting by the operator in the square brackets of Eq. (\ref{EOM2}) on Eqs. (\ref{EOM1}), the new EOM for the quasiparticle propagator is obtained as:
\bea
\left[i\partial_t \right. &-& \left.{\hat\sigma}_3E_{\mu\mu'}\right]{\hat{\cal R}}_{\mu\mu'\nu\nu'}(t-t') \left[-i\overleftarrow{\partial_{t'}} - {\hat\sigma}_3E_{\nu\nu'}\right] = \nonumber\\
&=& \delta(t-t'){\hat{\cal N}}_{\mu\mu'\nu\nu'}\left[-i\overleftarrow{\partial_{t'}} - {\hat\sigma}_3E_{\nu\nu'}\right] \nonumber\\
&+&  \delta(t-t'){\hat{\cal T}}^{0}_{\mu\mu'\nu\nu'} + {\hat{\cal T}}^{r}_{\mu\mu'\nu\nu'}(t-t'),
\label{EOMcomb}
\eea
where we have further denoted
\be
E_{\mu\mu'} = E_{\mu} + E_{\mu'}, \ \ \ \ \ \ \ \ \ \ \ \ \ \ \ {\hat\sigma}_3 = \left(\begin{array}{cc} 1& 0 \\ 0 & -1 \end{array}\right).
\ee 
At this step, it is convenient to make a Fourier transformation. The transformation of Eq. (\ref{EOMcomb}) to the energy (frequency) domain yields
\bea
\left[\omega\right. &-& \left.{\hat\sigma}_3E_{\mu\mu'}\right]{\hat{\cal R}}_{\mu\mu'\nu\nu'}(\omega) \left[\omega\right. - \left.{\hat\sigma}_3E_{\nu\nu'}\right] = \nonumber\\
&=& {\hat{\cal N}}_{\mu\mu'\nu\nu'}\left[\omega - {\hat\sigma}_3E_{\nu\nu'}\right] 
+  {\hat{\cal T}}^{0}_{\mu\mu'\nu\nu'} + {\hat{\cal T}}^{r}_{\mu\mu'\nu\nu'}(\omega). \nonumber\\
\label{EOMcomb_omega}
\eea
%with
%\bea
%{\hat{\cal R}}_{\mu\mu'\nu\nu'}(\omega) = \int\limits_{-\infty}^{\infty} d\tau e^{i\omega\tau}{\hat{\cal R}}_{\mu\mu'\nu\nu'}(\tau),
%\label{FI}\nonumber\\
%{\hat{\cal T}}^{r}_{\mu\mu'\nu\nu'}(\omega) = \int\limits_{-\infty}^{\infty} d\tau e^{i\omega\tau}{\hat{\cal T}}^{r}_{\mu\mu'\nu\nu'}(\tau).
%\eea
Then, after defining the free response as
\be
{\hat{\cal R}}^0_{\mu\mu'\nu\nu'}(\omega) = \left[\omega\right. - \left.{\hat\sigma}_3E_{\mu\mu'}\right]^{-1}{\hat{\cal N}}_{\mu\mu'\nu\nu'},
\ee
Eq. (\ref{EOMcomb_omega}) takes the form of a $T$-matrix equation:
\bea
{\hat{\cal R}}_{\mu\mu'\nu\nu'}(\omega) &=& {\hat{\cal R}}^0_{\mu\mu'\nu\nu'}(\omega) \nonumber\\
&+& 
\frac{1}{4}
{\hat{\cal R}}^0_{\mu\mu'\gamma\gamma'}(\omega){\hat{\cal T}}_{\gamma\gamma'\delta\delta'}(\omega){\hat{\cal R}}^0_{\delta\delta'\nu\nu'}(\omega)
\eea
with the energy-dependent $T$-matrix ${\hat{\cal T}}_{\gamma\gamma'\delta\delta'}(\omega)$ such, that
\be
{\hat{\cal T}}_{\gamma\gamma'\delta\delta'}(\omega) = \frac{1}{4}{\hat{\cal N}}^{-1}_{\gamma\gamma'\mu\mu'}
\bigl({\hat{\cal T}}^{0}_{\mu\mu'\nu\nu'} + {\hat{\cal T}}^{r}_{\mu\mu'\nu\nu'}(\omega)\bigr)
{\hat{\cal N}}^{-1}_{\nu\nu'\delta\delta'},
\ee
while we assume that the inverse norm matrix is defined according to the following identity:
\bea
\frac{1}{2}\sum_{\delta\delta'}{\hat{\cal N}}^{-1}_{\mu\mu'\delta\delta'}{\hat{\cal N}}_{\delta\delta'\nu\nu'} = \delta_{\mu\mu'\nu\nu'} =
 \delta_{\mu\nu}\delta_{\mu'\nu'} - \delta_{\mu\nu'}\delta_{\mu'\nu}.\nonumber\\
\eea
Again, as in the non-superfluid case, the $T$-matrix equation can be transformed to the Bethe-Salpeter-Dyson equation (BSDE) by  introducing the irreducible with respect to ${\hat{\cal R}}^0$ interaction kernel ${\hat{\cal K}}(\omega) = {\hat{\cal K}}^0 + {\hat{\cal K}}^r(\omega)$, which plays the role of the self-energy for the two-point two-fermion correlation function ${\hat{\cal R}}(\omega)$:
\bea
{\hat{\cal K}}^0_{\gamma\gamma'\delta\delta'} = \frac{1}{4}{\hat{\cal N}}^{-1}_{\gamma\gamma'\eta\eta'}
{\hat{\cal T}}^{0}_{\eta\eta'\rho\rho'} 
{\hat{\cal N}}^{-1}_{\rho\rho'\delta\delta'} \nonumber\\
{\hat{\cal K}}^r_{\gamma\gamma'\delta\delta'}(\omega) = \frac{1}{4}\left[{\hat{\cal N}}^{-1}_{\gamma\gamma'\eta\eta'}
{\hat{\cal T}}^{r}_{\eta\eta'\rho\rho'}(\omega) 
{\hat{\cal N}}^{-1}_{\rho\rho'\delta\delta'}\right]^{irr}.
\eea
Thus, the BSDE takes the familiar form
\bea
{\hat{\cal R}}_{\mu\mu'\nu\nu'}(\omega) = {\hat{\cal R}}^0_{\mu\mu'\nu\nu'}(\omega) \nonumber\\
+ \frac{1}{4}{\hat{\cal R}}^0_{\mu\mu'\gamma\gamma'}(\omega){\hat{\cal K}}_{\gamma\gamma'\delta\delta'}(\omega){\hat{\cal R}}_{\delta\delta'\nu\nu'}(\omega),
\label{BSDE}
\eea
but with the $2\times 2$ matrix structure in the quasiparticle basis.

\section{Interaction kernels}
The interaction kernels (\ref{T0},\ref{Tr}) require evaluation of commutators of the residual interaction $V$ of Eq. (\ref{Hqua2}) with the quasiparticle pair operators $A_{\mu\mu'}$ and $A^{\dagger}_{\mu\mu'}$ as well as the double commutators of Eq. (\ref{T0}) and the commutator products of Eq. (\ref{Tr}). The relevant generic commutators are given in Appendix \ref{AppB}, while further treatment of the kernels is presented in the next two subsections.

\subsection{The static kernel}

With respect to the static kernel, at this point it is useful to benchmark the obtained EOM in the BSDE form (\ref{BSDE}) to the well-established equation of the quasiparticle random phase approximation. The latter can be derived by (i) neglecting the irreducible part of the dynamical kernel ${\hat{\cal T}}^r$ (\ref{Tr})
and (ii) approximating the ground state wave function by the HFB wave function. Under these assumptions, Eq. (\ref{BSDE}) at the pole of the response function $\omega \to \omega_n$ takes the form \footnote{In this and the following subsection we indicate the summations over the quasiparticle states explicitly.}:
\bea
\left(\begin{array}{cc} {\cal X}^{n}_{\mu\mu'}{\cal X}^{n\ast}_{\nu\nu'}  &  {\cal X}^{n}_{\mu\mu'}{\cal Y}^{n\ast}_{\nu\nu'}
\\ \\
{\cal Y}^{n}_{\mu\mu'}{\cal X}^{n\ast}_{\nu\nu'}  &  {\cal Y}^{n}_{\mu\mu'}{\cal Y}^{n\ast}_{\nu\nu'}\end{array}\right) = 
\frac{1}{4}
\sum\limits_{\delta\delta',\gamma\gamma'}\left[\omega\right. - \left.{\hat\sigma}_3E_{\mu\mu'}\right]^{-1}\nonumber
\\
\times
\left(\begin{array}{cc} {\cal A}_{\mu\mu'\delta\delta'}-E_{\mu\mu'}\delta_{\mu\delta}\delta_{\mu'\delta'}&-{\cal B}_{\mu\mu'\delta\delta'}
\\ \\
-{\cal B}^{\ast}_{\mu\mu'\delta\delta'}&{\cal A}^{\ast}_{\mu\mu'\delta\delta'}-E_{\mu\mu'}\delta_{\mu\delta}\delta_{\mu'\delta'}\end{array}\right)
\nonumber\\
\times
{\hat{\cal N}}^{-1}_{\delta\delta'\gamma\gamma'}\left(\begin{array}{cc} {\cal X}^{n}_{\gamma\gamma'}{\cal X}^{n\ast}_{\nu\nu'}  &  {\cal X}^{n}_{\gamma\gamma'}{\cal Y}^{n\ast}_{\nu\nu'}
\\ \\
{\cal Y}^{n}_{\gamma\gamma'}{\cal X}^{n\ast}_{\nu\nu'}  &  {\cal Y}^{n}_{\gamma\gamma'}{\cal Y}^{n\ast}_{\nu\nu'}\end{array}\right),\nonumber\\
\label{toQRPA}
\eea
where
\bea
{\cal A}_{\mu\mu'\delta\delta'} &=& -\langle\text{HFB}|\left[[V,A_{\mu\mu'}],A^{\dagger}_{\delta\delta'}\right]|\text{HFB}\rangle 
+ E_{\mu\mu'}\delta_{\mu\delta}\delta_{\mu'\delta'},\nonumber
\\
{\cal B}_{\mu\mu'\delta\delta'} &=& \langle\text{HFB}|\left[[V,A_{\mu\mu'}],A_{\delta\delta'}\right]|\text{HFB}\rangle ,\nonumber
\\
{\hat{\cal N}}^{-1}_{\delta\delta'\gamma\gamma'} &=& {\hat{\sigma}}_3\delta_{\delta\delta'\gamma\gamma'}.
\nonumber\\
\label{ABN}
\eea
Eq. (\ref{toQRPA}) further simplifies to 
\bea
\left(\begin{array}{cc} {\cal X}^{n}_{\mu\mu'}{\cal X}^{n\ast}_{\nu\nu'}  &  {\cal X}^{n}_{\mu\mu'}{\cal Y}^{n\ast}_{\nu\nu'}
\\ \\
{\cal Y}^{n}_{\mu\mu'}{\cal X}^{n\ast}_{\nu\nu'}  &  {\cal Y}^{n}_{\mu\mu'}{\cal Y}^{n\ast}_{\nu\nu'}\end{array}\right) = 
\sum\limits_{\delta\leq\delta'}\nonumber
\\
\times
\Large{
\left(\begin{array}{cc} \frac{{\cal A}_{\mu\mu'\delta\delta'}-E_{\mu\mu'}\delta_{\mu\delta}\delta_{\mu'\delta'}}{\omega-E_{\mu\mu'}}&
\frac{{\cal B}_{\mu\mu'\delta\delta'}}{\omega-E_{\mu\mu'}}
\\
\\
-\frac{{\cal B}^{\ast}_{\mu\mu'\delta\delta'}}{\omega+E_{\mu\mu'}}&
\frac{-{\cal A}^{\ast}_{\mu\mu'\delta\delta'}+E_{\mu\mu'}\delta_{\mu\delta}\delta_{\mu'\delta'}}{\omega+E_{\mu\mu'}}\end{array}\right)
}
\nonumber\\
\times
\left(\begin{array}{cc} {\cal X}^{n}_{\delta\delta'}{\cal X}^{n\ast}_{\nu\nu'}  &  {\cal X}^{n}_{\delta\delta'}{\cal Y}^{n\ast}_{\nu\nu'}
\\ \\
{\cal Y}^{n}_{\delta\delta'}{\cal X}^{n\ast}_{\nu\nu'}  &  {\cal Y}^{n}_{\delta\delta'}{\cal Y}^{n\ast}_{\nu\nu'}\end{array}\right),\nonumber\\
\label{toQRPA_1}
\eea
%with $\mu\leq\mu', \nu\leq\nu'$, 
that, after eliminating the obvious redundancies, yields the QRPA equations for the ${\cal X}^{n}$ and ${\cal Y}^{n}$ amplitudes in the conventional form:
\bea
\omega{\cal X}^{n}_{\mu\mu'} = \sum\limits_{\delta\leq\delta'}({\cal A}_{\mu\mu'\delta\delta'}{\cal X}^{n}_{\delta\delta'} +
{\cal B}_{\mu\mu'\delta\delta'}{\cal Y}^{n}_{\delta\delta'})\nonumber\\
\omega{\cal Y}^{n}_{\mu\mu'} = \sum\limits_{\delta\leq\delta'}(-{\cal B}^{\ast}_{\mu\mu'\delta\delta'}{\cal X}^{n}_{\delta\delta'} -
{\cal A}^{\ast}_{\mu\mu'\delta\delta'}{\cal Y}^{n}_{\delta\delta'})
\label{QRPA}
\eea
or, in terms of block matrices,
\be
\left(\begin{array}{cc} {\cal A} & {\cal B} \\ -{\cal B}^{\ast} & -{\cal A}^{\ast}\end{array}\right)
\left(\begin{array}{c} {\cal X}^n\\{\cal Y}^n\end{array}\right) = \omega \left(\begin{array}{c} {\cal X}^n\\{\cal Y}^n\end{array}\right).
\ee

The explicit forms of the ${\cal A}$ and ${\cal B}$ matrices, from Eq. (\ref{ABN}), are
\bea
{\cal A}_{\mu\mu'\nu\nu'} &=& E_{\mu\mu'}\delta_{\mu\nu}\delta_{\mu'\nu'} + H^{22}_{\mu\mu'\nu\nu'}\\
{\cal B}_{\mu\mu'\nu\nu'} &=& 4!H^{40}_{\mu\mu'\nu\nu'},
\label{QRPA_mat}
\eea
that follows from the explicit evaluation of the double commutators of Eq. (\ref{T0}).
% This was incorrect. The matrices are the same, after the cross-check.

The exact form of the static kernel in the absence of pairing correlations was presented and discussed, e.g., in Refs. \cite{SchuckTohyama2016a,LitvinovaSchuck2019,Schuck2021}. Besides the pure contribution from the bare fermionic interaction, it contains the terms with contractions of the interaction with the correlated parts of the two-body fermionic densities which include, in principle, a feedback from the dynamical kernel. The superfluid analogs of these terms will be discussed elsewhere.

Another remark pertains to the norm matrix defined by Eq. (\ref{norm}), which should have a non-divergent inverse in order for the whole approach to be well defined. This requirement is fulfilled in the HF(B) approximation implied in this work, however, difficulties may occur in the case of correlated ground states. As it has been pointed out in Ref. \cite{DukelskyRoepkeSchuck1998}, the norm matrix must have non-zero eigenvalues, otherwise the approach generates spurious components, which should be eliminated. An example of the latter procedure is given by the generator coordinate method discussed, e.g., in Ref. \cite{RingSchuck1980}.

\subsection{The dynamical kernel}

The dynamical kernel defined by the time-dependent commutator products of Eq. (\ref{Tr}) can be evaluated component by component. The upper left component reads
\be
{\cal T}^{r[11]}_{\mu\mu'\nu\nu'}(t-t') = i\langle T[V,A_{\mu\mu'}](t)[V,A^{\dagger}_{\nu\nu'}](t') 
\rangle
\label{Tr11}
\ee
with the commutators given by Eqs. (\ref{VAcomm}, \ref{VAdcomm}). It is straightforward to see that each term in Eq. (\ref{Tr11}) contains a product of eight quasiparticle operators, four at time $t$ and four at time $t'$, i.e., fully correlated two-times four-quasiparticle propagator, contracted with two matrix elements of the residual interaction. As in the non-superfluid case, the appearance of higher-rank propagators in the dynamical kernel signals generating a hierarchy of coupled EOMs for growing-rank propagators.  Approximations have to be applied to treat such dynamical kernels, and various approximations of increasing accuracy constructed by a factorization procedure are possible, some of which were discussed, for instance, in Ref. \cite{LitvinovaSchuck2019} for the non-superfluid case. 

The simplest approximation can be obtained by the complete factorization of the correlated four-quasiparticle propagator into one-quasiparticle ones, that would be the superfluid analog of the second random phase approximation \cite{YannouleasJangChomaz1985}. In this work we will, however, focus on the other types of factorization, which lead to more accurate approximations because of retaining important effects of emergent collectivity. These effects are especially significant in the strong-coupling regimes.   
Namely, we will (i) perform factorizations into pairwise products of the fully correlated two-quasiparticle propagators and then (ii) relax the correlations in one correlation function of each pair. The latter approach will be, thus, the superfluid version of the conventional particle-vibration coupling dynamical kernel, which we will address as quasiparticle-vibration coupling (qPVC), and the former one will correspond to the superfluid two-phonon model. We note here that the phonons appearing in these factorizations are formally exact and, in general, are not associated with any partial resummations or perturbative expansions. This means that such approaches can include, in principle, arbitrarily complex 2n-quasiparticle configurations of the certain kinds. However, approximations can always be applied to calculations of these phonons.

Although the form of the two-quasiparticle response function (\ref{Rt}) is analogous to the one of the pair propagator discussed, for instance, in Ref. \cite{LitvinovaSchuck2020}, the residual interaction (\ref{Hqua2}) has now a more complicated structure. It has been established, in particular, in Ref. \cite{RingSchuck1980} that the terms associated with $H^{40}$ are responsible for the complex ground state correlations, while the $H^{22}$ contributions have the same operator structure. The leading contributions, thus, come from the $H^{31}$ terms. As one can see from Eq. (\ref{VAcomm}), there are three types of operator products at $H^{31}$: (i) those containing only $A$ and $A^{\dagger}$ operators, (ii) those with only $C_{\mu\nu} = \alpha^{\dagger}_{\mu}\alpha_{\nu}$ operators and (iii) products of $A$ ($A^{\dagger}$) and $C$ operators.

Let us consider the first group of terms ('$AA$'). Their contribution to ${\cal T}^{r[11]}_{\mu\mu'\nu\nu'}(\tau)$ reads:
\bea
{\cal T}^{r[11]AA}_{\mu\mu'\nu\nu'}(\tau) = -i\langle T\sum\limits_{\rho\rho'\gamma} \Bigl[\bigl(H^{31}_{\rho\rho'\mu\gamma}(A^{\dagger}_{\rho\rho'}A_{\mu'\gamma})(t) \nonumber \\ + 
H^{31\ast}_{\rho\rho'\gamma\mu}(A_{\mu'\gamma}A_{\rho\rho'})(t)\bigr) - (\mu\leftrightarrow\mu')\Bigr]\nonumber \\ 
\times \sum\limits_{\eta\eta'\delta} \Bigl[\bigl(H^{31\ast}_{\eta\eta'\nu\delta}(A^{\dagger}_{\nu'\delta}A_{\eta\eta'})(t') \nonumber \\ + 
H^{31}_{\eta\eta'\delta\nu}(A^{\dagger}_{\eta\eta'}A^{\dagger}_{\nu'\delta})(t')\bigr) - (\nu\leftrightarrow\nu')\Bigr] \rangle,
\nonumber \\ 
\eea
where $\tau = t-t'$. The products of the terms listed explicitly admit a number of irreducible factorizations. Let us consider those without breaking the two-quasiparticle pairs in the operators $A_{\mu\mu'}$ and $A^{\dagger}_{\mu\mu'}$:
\bea
\langle T (A^{\dagger}_{\rho\rho'}A_{\mu'\gamma})(t)(A^{\dagger}_{\nu'\delta}A_{\eta\eta'})(t')\rangle \nonumber\\
\approx 
\langle T A^{\dagger}_{\rho\rho'}(t)A^{\dagger}_{\nu'\delta}(t')\rangle\langle T A_{\mu'\gamma}(t)A_{\eta\eta'}(t')\rangle \nonumber \\ +
\langle T A^{\dagger}_{\rho\rho'}(t)A_{\eta\eta'}(t')\rangle\langle T A_{\mu'\gamma}(t)A^{\dagger}_{\nu'\delta}(t')\rangle \nonumber \\ 
= -R^{[21]}_{\rho\rho'\nu'\delta}(\tau)R^{[12]}_{\mu'\gamma\eta\eta'}(\tau) - R^{[22]}_{\rho\rho'\eta\eta'}(\tau)R^{[11]}_{\mu'\gamma\nu'\delta}(\tau), \nonumber\\
\label{Tr11AA1}
\\
\langle T (A^{\dagger}_{\rho\rho'}A_{\mu'\gamma})(t)(A^{\dagger}_{\eta\eta'}A^{\dagger}_{\nu'\delta})(t')\rangle \nonumber\\
\approx 
\langle T A^{\dagger}_{\rho\rho'}(t)A^{\dagger}_{\eta\eta'}(t')\rangle\langle T A_{\mu'\gamma}(t)A^{\dagger}_{\nu'\delta}(t')\rangle \nonumber \\ +
\langle T A^{\dagger}_{\rho\rho'}(t)A^{\dagger}_{\nu'\delta}(t')\rangle\langle T A_{\mu'\gamma}(t)A^{\dagger}_{\eta\eta'}(t')\rangle \nonumber \\ 
= -R^{[21]}_{\rho\rho'\eta\eta'}(\tau)R^{[11]}_{\mu'\gamma\nu'\delta}(\tau) - R^{[21]}_{\rho\rho'\nu'\delta}(\tau)R^{[11]}_{\mu'\gamma\eta\eta'}(\tau), \nonumber\\
\label{Tr11AA2}
\\
\langle T (A_{\mu'\gamma}A_{\rho\rho'})(t)(A^{\dagger}_{\nu'\delta}A_{\eta\eta'})(t')\rangle \nonumber\\
\approx 
\langle T A_{\mu'\gamma}(t)A^{\dagger}_{\nu'\delta}(t')\rangle\langle T A_{\rho\rho'}(t)A_{\eta\eta'}(t')\rangle \nonumber \\ +
\langle T A_{\mu'\gamma}(t)A_{\eta\eta'}(t')\rangle \langle T A_{\rho\rho'}(t)A^{\dagger}_{\nu'\delta}(t')\rangle\nonumber \\ 
= -R^{[11]}_{\mu'\gamma\nu'\delta}(\tau)R^{[12]}_{\rho\rho'\eta\eta'}(\tau) - R^{[12]}_{\mu'\gamma\eta\eta'}(\tau)R^{[11]}_{\rho\rho'\nu'\delta}(\tau), \nonumber\\
\label{Tr11AA3}
\\
\langle T (A_{\mu'\gamma}A_{\rho\rho'})(t)(A^{\dagger}_{\eta\eta'}A^{\dagger}_{\nu'\delta})(t')\rangle \nonumber\\
\approx 
\langle T A_{\mu'\gamma}(t)A^{\dagger}_{\eta\eta'}(t')\rangle\langle T A_{\rho\rho'}(t)A^{\dagger}_{\nu'\delta}(t')\rangle \nonumber \\ +
\langle T A_{\mu'\gamma}(t)A^{\dagger}_{\nu'\delta}(t')\rangle \langle T A_{\rho\rho'}(t)A^{\dagger}_{\eta\eta'}(t')\rangle\nonumber \\ 
= -R^{[11]}_{\mu'\gamma\eta\eta'}(\tau)R^{[11]}_{\rho\rho'\nu'\delta}(\tau) - R^{[11]}_{\mu'\gamma\nu'\delta}(\tau)R^{[11]}_{\rho\rho'\eta\eta'}(\tau). \nonumber\\
\label{Tr11AA4}
\eea
The corresponding contribution to ${\cal K}^{r[11]AA}$ reads:
\bea
{\cal K}^{r[11]AA}_{\mu\mu'\nu\nu'}(\tau) \approx 
\Bigl\{\Bigl[
i\sum\limits_{\rho\rho'\gamma}\sum\limits_{\eta\eta'\delta}\bigl[H^{31}_{\rho\rho'\mu\gamma}H^{31\ast}_{\eta\eta'\nu\delta}
\nonumber\\
\times\bigl( R^{[22]}_{\rho\rho'\eta\eta'}(\tau)R^{[11]}_{\mu'\gamma\nu'\delta}(\tau) + R^{[21]}_{\rho\rho'\nu'\delta}(\tau)R^{[12]}_{\mu'\gamma\eta\eta'}(\tau)\bigr)
\nonumber\\
+ H^{31}_{\rho\rho'\mu\gamma}H^{31}_{\eta\eta'\delta\nu}
\nonumber\\
\times\bigl( R^{[21]}_{\rho\rho'\eta\eta'}(\tau)R^{[11]}_{\mu'\gamma\nu'\delta}(\tau) + R^{[21]}_{\rho\rho'\nu'\delta}(\tau)R^{[11]}_{\mu'\gamma\eta\eta'}(\tau)\bigr)
\nonumber\\
+ H^{31\ast}_{\rho\rho'\gamma\mu}H^{31\ast}_{\eta\eta'\nu\delta}
\nonumber\\
\times\bigl( R^{[12]}_{\rho\rho'\eta\eta'}(\tau)R^{[11]}_{\mu'\gamma\nu'\delta}(\tau) + R^{[11]}_{\rho\rho'\nu'\delta}(\tau)R^{[12]}_{\mu'\gamma\eta\eta'}(\tau)\bigr)
\nonumber\\
+ H^{31\ast}_{\rho\rho'\gamma\mu}H^{31}_{\eta\eta'\delta\nu}
\nonumber\\
\times\bigl( R^{[11]}_{\rho\rho'\eta\eta'}(\tau)R^{[11]}_{\mu'\gamma\nu'\delta}(\tau) + R^{[11]}_{\rho\rho'\nu'\delta}(\tau)R^{[11]}_{\mu'\gamma\eta\eta'}(\tau)\bigr)
\bigr]\Bigr]  \nonumber\\ 
- \Bigl[ \mu\leftrightarrow\mu'\Bigr]\Bigr\} - \Bigl\{ \nu\leftrightarrow\nu'\Bigr\}.
\nonumber\\
\label{Kr11AA}
\eea

At this point it may be instructive to use the explicit form of the components of the superfluid response $R^{[ij]}$, in particular, the fact that the residues of each component factorize with respect to the first and second pairs of the quasiparticle indices (\ref{R}). Therefore, it is convenient to perform the Fourier transformation of the dynamical kernel to the frequency domain. For the generic product of two components like those figuring in Eq. (\ref{Kr11AA}), we have:
\bea
[R^{[ij]}_{\mu\mu'\nu\nu'}R^{[kl]}_{\eta\eta'\rho\rho'}](\omega) = \int_{-\infty}^{\infty}d\tau e^{i\omega\tau}R^{[ij]}_{\mu\mu'\nu\nu'}(\tau)R^{[kl]}_{\eta\eta'\rho\rho'}(\tau) \nonumber\\
= -i\sum\limits_{nm}\sum\limits_{\sigma = \pm}\sigma\frac{{\cal Z}^{n(i\sigma)}_{\mu\mu'}{\cal Z}^{n(j\sigma)\ast}_{\nu\nu'}
{\cal Z}^{m(k\sigma)}_{\eta\eta'}{\cal Z}^{m(l\sigma)\ast}_{\rho\rho'}}
{\omega - \sigma(\omega_{nm}-i\delta)},
\label{FT}
\nonumber\\
\eea
where
\bea
{\cal Z}^{n(1+)}_{\mu\mu'} = {\cal X}^{n}_{\mu\mu'}, \ \ \ \ \ \ \ \ \ \ \ \ \ {\cal Z}^{n(2+)}_{\mu\mu'} = {\cal Y}^{n}_{\mu\mu'}, \nonumber\\
{\cal Z}^{n(1-)}_{\mu\mu'} = {\cal Y}^{n\ast}_{\mu\mu'}, \ \ \ \ \ \ \ \ \ \ \ \ \ {\cal Z}^{n(2-)}_{\mu\mu'} = {\cal X}^{n\ast}_{\mu\mu'},
\eea
the upper indices $i,k,l,m = \{1,2\}$, and $\omega_{nm} = \omega_{n} + \omega_{m}$.
\begin{figure}
\begin{center}
%\vspace{-0.3cm}
\includegraphics[scale=0.52]{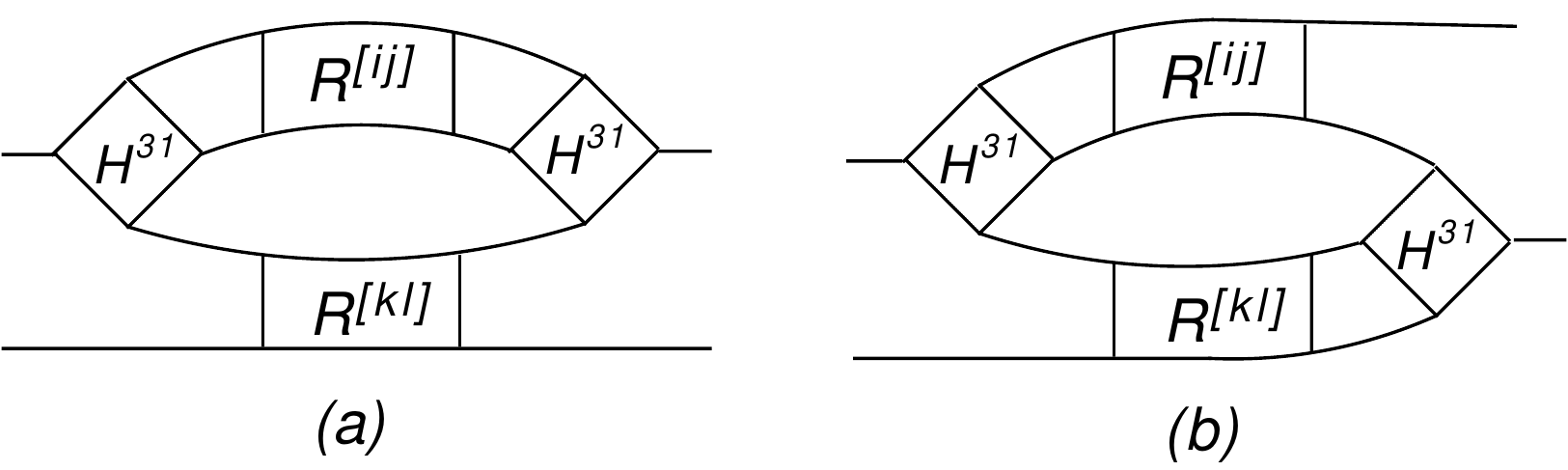}
\end{center}
%\vspace{-6.3cm}
\caption{Two types of the leading ($H^{31}$-associated) irreducible contributions to the dynamical kernel ${\cal K}^r$ after factorization into the products of two two-quasiparticle correlation functions.}
\label{Kr}%
\end{figure}

Furthermore, analyzing the structure of the contributions to ${\cal K}^{r[11]AA}$, one can separate them into two groups. They are represented by the first and the second terms in the round brackets of Eq. (\ref{Kr11AA}) and can be associated with the graphs (a) and (b) in Fig. \ref{Kr}, respectively. As we will see in the following, all the remaining $H^{31}$-associated contributions to ${\cal K}^{r[11]}$ will also belong to these two classes. 
The graphs of type (a) are of the special interest as they have the topological structure analogous to that appearing in the non-superfluid theory of the particle-vibration coupling, see for instance, Ref. \cite{LitvinovaSchuck2019}. Group (b) is associated with a rearrangement of correlations between the two two-quasiparticle 
pairs of the four-quasiparticle propagator and represents a further extension of the qPVC approach. The analogous cluster decomposition and a similar grouping can be obtained for the terms of the '$CC$' and '$AC$' types, which are given in appendix \ref{AppC}, processed along with the '$AA$' terms.

%Thus, after applying the (exact) mapping introduced in Ref. \cite{LitvinovaSchuck2019},
Summing up all the terms, 
the total contribution of type (a) to the dynamical kernel in the energy domain can be recast as follows:
%%%%%%%%%%%%%%
% Moved to Appendix C
%%%%%%%%%%%%%%
\bea
{\cal K}^{r[11]cc}_{\mu\mu'\nu\nu'}(\omega) &=& 
%\frac{1}{4}
%\Bigl\{\Bigl[
\sum\limits_{\gamma\delta nm}\Bigl[\frac{\Gamma^{(11)n}_{\mu\gamma}{\cal X}^{m}_{\mu'\gamma}{\cal X}^{m\ast}_{\nu'\delta}\Gamma^{(11)n\ast}_{\nu\delta}}{\omega - \omega_{nm} + i\delta} \nonumber\\ &-& 
\frac{\Gamma^{(11)n\ast}_{\gamma\mu}{\cal Y}^{m\ast}_{\mu'\gamma}{\cal Y}^{m}_{\nu'\delta}\Gamma^{(11)n}_{\delta\nu}}{\omega + \omega_{nm} - i\delta}\Bigr]
%\Bigr]
- \cal{AS},
\label{Kr11cc}
\eea
where the "combined" vertices $\Gamma^{(11)n}$ are introduced as 
\bea
\Gamma^{(11)n}_{\mu\gamma} = \frac{1}{2}\bigl(\theta^n_{\mu\gamma} + \xi^n_{\mu\gamma} + {\bar\xi}^n_{\mu\gamma}\bigr) \nonumber\\
= 
\Bigl[ 
U^{\dagger}g^{n}U + U^{\dagger}\gamma^{n(+)}V 
- V^{\dagger}g^{nT}V - V^{\dagger}\gamma^{n(-)T}U\Bigr]_{\mu\gamma}, \nonumber\\
\label{Gamma11_HFB}
\eea
with the "partial" vertices $\theta^n$, $\xi^n$ and ${\bar\xi}^n$ defined in Eqs. (\ref{vert1},\ref{vert23}).
The additional index "cc" in Eq. (\ref{Kr11cc}) is used to mark the approach with two correlated two-quasiparticle propagators in the dynamical kernel.
In Eqs. (\ref{Kr11cc}, \ref{Gamma11_HFB}) we employed (i) the explicit definition of $H^{31}$ (\ref{H31}), (ii) the notions of the normal and anomalous transition densities 
\bea
\rho^{n}_{12} &=& \langle 0|\psi^{\dagger}_2\psi_1|n \rangle \nonumber\\
\varkappa^{n(+)}_{12} &=& \langle 0|\psi_2\psi_1|n\rangle \nonumber\\
 \varkappa^{n(-)\ast}_{21} &=& \langle 0|\psi^{\dagger}_2\psi^{\dagger}_1|n\rangle ,
\label{rho}
\eea
in connection to ${\cal X}^n$ and ${\cal Y}^n$ amplitudes, that follows from the Bogolyubov's transformation (\ref{Btrans})
\bea
\rho^n_{12} &=&  (U{\cal X}^nV^T + V^{\ast}{\cal Y}^{nT}U^{\dagger})_{12}\nonumber\\
\varkappa^{n(+)}_{12} &=& (U{\cal X}^nU^T + V^{\ast}{\cal Y}^{nT}V^{\dagger})_{12}\nonumber\\
\varkappa^{n(-)}_{12} &=& (V^{\ast}{\cal X}^{n\dagger}V^{\dagger} + U{\cal Y}^{n\ast}U^T)_{12},
\label{Dens}
\eea
in the approximation of vanishing $\delta{\cal R}^{11}$ components (see the definition of $\delta{\cal R}$ in the next section),
(iii) the definition of the normal and pairing phonon vertices, respectively \cite{LitvinovaSchuck2019,Litvinova2021},
\bea
g^{n}_{13} = \sum\limits_{24}{\bar v}_{1234}\rho^{n}_{42}, \ \ \ \ \ \ \ \ \ \ \ \ \ \ \ \ \ \ \nonumber\\
\gamma^{n(+)}_{12} = \frac{1}{2}\sum\limits_{34} {\bar v}_{1234}\varkappa^{n(+)}_{34}, \ \ \ \ \ \ \gamma_{12}^{n(-)T} = \frac{1}{2}\sum\limits_{34}{\bar v}^{\ast}_{1234}\varkappa^{n(-)\ast}_{34}, \nonumber \\
\label{Vert}
\eea
and (iv) the obvious properties of the antisymmetrized bare interaction: ${\bar v}_{1234} = -{\bar v}_{2134} = -{\bar v}_{1243} = {\bar v}^{\ast}_{3412}$.

As one can see from Eq. (\ref{Gamma11_HFB}), the obtained combined vertex has the familiar structure of the (11)-component of a single-particle operator in the quasiparticle space \cite{RingSchuck1980}. This fact is reflected by the upper index "(11)". This vertex was obtained in our previous work \cite{Litvinova2021} devoted to the EOM for a single-quasiparticle propagator and, in particular, to its dynamical kernel. The vertex $\Gamma^{(11)n}$ enters the forward-going component of the dynamical qPVC self-energy, while its counterpart $\Gamma^{(02)n}$ was found to be responsible for the backward-going component, i.e., the ground state correlations induced by the qPVC. As expected, the latter vertex does not appear in the kernel of Eq. (\ref{Kr11cc}) because it does not include the qPVC-induced ground state correlations GSC-qPVC. These correlations may, in some cases, play a non-negligible role, however, taking them into account is a difficult task, which was up until now implemented  
only rarely and only to non-superfluid phases \cite{KamerdzhievSpethTertychny2004,Drozdz:1990zz,Robin2019,Litvinova2021a}, although a superfluid version of GSC-qPVC was formulated in Ref. \cite{Tselyaev2007} for the BCS-type superfluidity. The full HFB-type superfluid GSC-qPVC can be derived by processing explicitly the corresponding terms dropped in this work and will be discussed elsewhere.

%emergent superfluidity 
%
\begin{figure*}
\begin{center}
%\vspace{-0.3cm}
\includegraphics[scale=0.70]{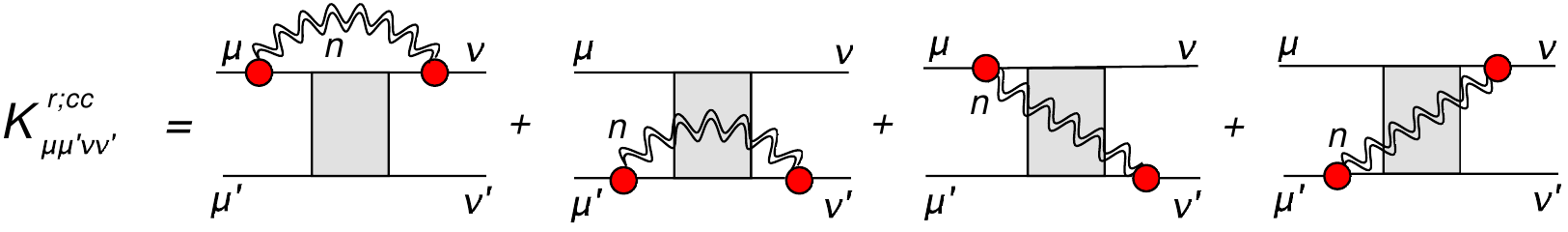}
\\
\vspace{0.5cm}
\includegraphics[scale=0.34]{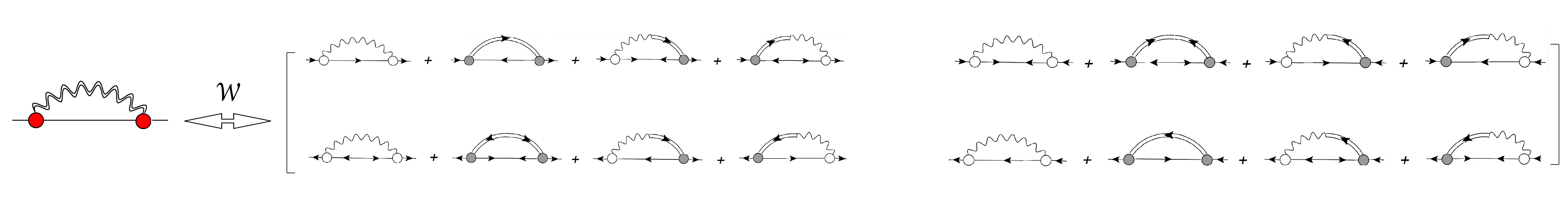}
\end{center}
%\vspace{-6.3cm}
\caption{{\it Top:} The leading approximation to the superfluid (qPVC) dynamical kernel. {\it Bottom:} the superfluid qPVC self-energy, where $\cal W$ stands for the Bogolyubov's transformation. Empty and filled (grey) circles in the self-energy matrix denote the normal and phonon vertices, while their  propagators are represented by the wavy and double lines, respectively. Single lines with arrows stand for fermionic particles (right arrows) and holes (left arrows), and single lines without arrows correspond to the quasiparticle propagators. Double wavy lines are reserved for the propagators of the superfluid phonons of the "unified" character in the quasiparticle basis, while the associated filled (red) circles stand for the respective combined phonon vertices. The rectangular block is used to represent the two-quasiparticle correlation function.}
\label{Kr_qvc}%
\end{figure*}
%

%%%%%%%%%Type (b) summary

The graphs of type (b) can be recast in a similar compact form:
\bea
{\tilde{\cal K}}^{r[11]cc}_{\mu\mu'\nu\nu'}(\omega) &=& 
%\frac{1}{4}
%\Bigl\{\Bigl[
\sum\limits_{\gamma\delta nm}\Bigl[\frac{\Gamma^{(11)n}_{\mu\gamma}{\cal X}^{m}_{\mu'\gamma}{\cal X}^{n\ast}_{\nu'\delta}\Gamma^{(11)m\ast}_{\nu\delta}}{\omega - \omega_{nm} + i\delta} \nonumber\\ &-& 
\frac{\Gamma^{(11)n\ast}_{\gamma\mu}{\cal Y}^{m\ast}_{\mu'\gamma}{\cal Y}^{n}_{\nu'\delta}\Gamma^{(11)m}_{\delta\nu}}{\omega + \omega_{nm} - i\delta}\Bigr]
%\Bigr]
- \cal{AS}
%\Bigl[ \mu\leftrightarrow\mu'\Bigr]\Bigr\} - \Bigl\{ \nu\leftrightarrow\nu'\Bigr\}
%\nonumber\\
\label{Kr11ccb}
\eea
and, thereby, lead to a further extension of the dynamical kernel.
The diagrammatic representation of Fig. \ref{Kr} indicates that the formation of the superfluid phonons is mainly associated with the type (a) contributions, while the type (b) contributions may be less important in the strong-coupling regime. This can be clarified by concrete calculations for realistic strongly-correlated systems and for exactly solvable Hamiltonians.
In the following we will concentrate on the type (a) contributions, while the ones of type (b) can be included in a similar manner.

Now we show how the superfluid generalization of the conventional NFT kernel can be obtained from the doubly-correlated kernel ${\cal K}^{r[11]cc}$.
As it was discussed in Ref. \cite{LitvinovaSchuck2019} for the normal phase, all the terms of such a kernel contain only one two-fermion correlation function. Thus, to obtain its superfluid analog, we can relax correlations in one of the two two-quasiparticle propagators. Technically, it means the following substitution of the correlated two-quasiparticle propagator by the uncorrelated one to be made in Eq. (\ref{Kr11cc}): $R^{[11]}_{\mu'\gamma\nu'\delta}(\omega) \to R^{0[11]}_{\mu'\gamma\nu'\delta}(\omega)$, where
\bea
R^{[11]}_{\mu'\gamma\nu'\delta}(\omega) &=& \sum\limits_{m}\Bigl(\frac{{\cal X}^m_{\mu'\gamma}{\cal X}^{m\ast}_{\nu'\delta}}{\omega - \omega_m} -
\frac{{\cal Y}^{m\ast}_{\mu'\gamma}{\cal Y}^{m}_{\nu'\delta}}{\omega + \omega_m}\Bigr), \\
R^{0[11]}_{\mu'\gamma\nu'\delta}(\omega) &=& \frac{\delta_{\mu'\nu'}\delta_{\gamma\delta} - \delta_{\mu'\delta}\delta_{\gamma\nu'}}{\omega - E_{\mu'} - E_{\gamma}},
\label{R011}
\eea
and where we dropped the infinitesimal imaginary parts of the energy variable because of their irrelevance. This procedure can be performed directly with the kernel of Eq. (\ref{Kr11cc}) as (i) ${\cal X}^m_{\mu'\gamma}{\cal X}^{m\ast}_{\nu'\delta} \to \delta_{\mu'\nu'}\delta_{\gamma\delta} - \delta_{\mu'\delta}\delta_{\gamma\nu'}$, (ii) $\omega_m \to E_{\mu'} + E_{\gamma}$, (iii) ${\cal Y}^{m\ast}_{\mu'\gamma}{\cal Y}^{m}_{\nu'\delta} \to 0$. Here we note that the formally infinite sum over $m$ reduces to one term as the phonon frequency $\omega_n$ is replaced by the summed energies of two quasiparticles $E_{\mu'} + E_{\gamma}$, and the second negative-frequency term vanishes, i.e., we drop also the ground state correlations of the two-phonon coupling. 

For the kernel (\ref{Kr11cc}) we, thus, obtain its qPVC-NFT approximation:
\bea
{\cal K}^{r[11]c}_{\mu\mu'\nu\nu'}(\omega) = 
%\frac{1}{4}
\Bigl\{\Bigl[ \delta_{\mu'\nu'}
\sum\limits_{\gamma n}\frac{\Gamma^{(11)n}_{\mu\gamma}\Gamma^{(11)n\ast}_{\nu\gamma}}{\omega - \omega_{n} - E_{\mu'} - E_{\gamma}}  \nonumber\\ - 
\sum\limits_{n}\frac{\Gamma^{(11)n}_{\mu\nu'}\Gamma^{(11)n\ast}_{\nu\mu'}}{\omega - \omega_{n} - E_{\mu'} - E_{\nu'}}\Bigr]
- \Bigl[ \mu\leftrightarrow\mu'\Bigr]\Bigr\} - \Bigl\{ \nu\leftrightarrow\nu'\Bigr\},\nonumber\\
%\cal{AS},
\label{Kr11AAa_qPVC_0}
\eea
where we indicated by the index "c" that only one two-quasiparticle correlation function is retained in the dynamical kernel.
To bring this kernel to the conventional NFT form, one can recast Eq. (\ref{Kr11AAa_qPVC_0}) by performing the explicit antisymmetrizations and rearranging the resulting terms as follows:
\bea
{\cal K}^{r[11]c}_{\mu\mu'\nu\nu'}(\omega) =  \ \ \ \ \ \ \ \ \ \ \ \ \ \ \ \ \ \ \ \  \nonumber\\
%\frac{1}{4} \nonumber\\
%\times
= \Bigl[ \delta_{\mu'\nu'}
\sum\limits_{\gamma n}\frac{\Gamma^{(11)n}_{\mu\gamma}\Gamma^{(11)n\ast}_{\nu\gamma}}{\omega - \omega_{n} - E_{\mu'\gamma}}  
+\delta_{\mu\nu}
\sum\limits_{\gamma n}\frac{\Gamma^{(11)n}_{\mu'\gamma}\Gamma^{(11)n\ast}_{\nu'\gamma}}{\omega - \omega_{n} - E_{\mu\gamma}}
\nonumber\\ + 
\sum\limits_{n}\frac{\Gamma^{(11)n}_{\mu\nu}\Gamma^{(11)n\ast}_{\nu'\mu'}}{\omega - \omega_{n} - E_{\mu'\nu}}
+ \sum\limits_{n}\frac{\Gamma^{(11)n}_{\mu'\nu'}\Gamma^{(11)n\ast}_{\nu\mu}}{\omega - \omega_{n} - E_{\mu\nu'}}\Bigr] \nonumber\\
- \Bigl[ \nu\leftrightarrow\nu'\Bigr].\ \ \ \ \ \ \ \ \ \ \ \ 
%\cal{AS}
\label{Kr11c}
\eea
As the interaction kernel is contracted with the antisymmetric quantities, such as ${\cal X}^n_{\nu\nu'}, {\cal Y}^n_{\nu\nu'}$, the antisymmetrization $[\nu\leftrightarrow\nu']$ is equivalent to the doubling of the corresponding terms. In practice, this double counting is taken care of by the numerical factor in Eq. (\ref{BSDE}), and further the restricted summation is implied over $\nu < \nu'$ as, for instance, in the QRPA equations.
%Thus, this antisymmetrization can be replaced by a factor of two, that will lead to the factor one half instead of one fourth in Eq. (\ref{Kr11AAa_qPVC}). %Furthermore, restricting the summations by $\nu < \nu'$ should remove the remaining factor one half from the final EOM.

The diagrammatic representation of the qPVC kernels of Eqs. (\ref{Kr11cc},\ref{Kr11c}) is given in the top part of Fig. \ref{Kr_qvc}, where the latter version differs from the former one by the absence of the two-quasiparticle correlations (rectangular blocks) between the emission and absorption of a phonon.
The terms shown explicitly correspond to those of Eq. (\ref{Kr11c}) in the same order. The first two contributions contain the two possible arrangements of the self-energy, while the third and the fourth ones are associated with the phonon exchange. The bottom part of Fig. \ref{Kr_qvc} illustrates the component structure of the superfluid phonons, which form the quasiparticle self-energy in the qPVC approximation \cite{Litvinova2021}. 
%higher symmetry than in normal phase.

The other components ${\cal K}^{r[12]}, {\cal K}^{r[21]}$ and ${\cal K}^{r[22]}$ of the dynamical kernel can be evaluated in a similar manner. From Eqs. (\ref{dK20}) and (\ref{dK02}) below one can see that the off-diagonal matrix elements of the dynamical kernel couple to the backward-going components of the density variations. It means that they represent the qPVC-associated ground-state correlations and can be dropped in the leading approximation. In this case, the only remaining component is ${\cal K}^{r[22]}$. From the definition of the dynamical T-matrix (\ref{Tr}), it is related to ${\cal K}^{r[11]}$ as follows:
\be
{\cal K}^{r[22]}_{\mu\mu'\nu\nu'}(\tau) = {\cal K}^{r[11]}_{\nu\nu'\mu\mu'}(-\tau).
\ee
%i.e., explicitly,
%\bea
%{\cal K}^{r[22]\text{qPVC}}_{\mu\mu'\nu\nu'}(\omega) = \frac{1}{4} \nonumber\\
%\times\Bigl[ \delta_{\mu'\nu'}
%\sum\limits_{\gamma n}\frac{\Gamma^{(11)n}_{\nu\gamma}\Gamma^{(11)n\ast}_{\mu\gamma}}{\omega + \omega_{n} + E_{\nu'\gamma}}  
%+\delta_{\mu\nu}
%\sum\limits_{\gamma n}\frac{\Gamma^{(11)n}_{\nu'\gamma}\Gamma^{(11)n\ast}_{\mu'\gamma}}{\omega + \omega_{n} + E_{\nu\gamma}}
%\nonumber\\ + 
%\sum\limits_{n}\frac{\Gamma^{(11)n}_{\nu\mu}\Gamma^{(11)n\ast}_{\mu'\nu'}}{\omega + \omega_{n} + E_{\nu'\mu}}
%+ \sum\limits_{n}\frac{\Gamma^{(11)n}_{\nu'\mu'}\Gamma^{(11)n\ast}_{\mu\nu}}{\omega + \omega_{n} + E_{\nu\mu'}}\Bigr] \nonumber\\
%- \Bigl[ \nu\leftrightarrow\nu'\Bigr].\ \ \ \ \ \ \ \ \ \ \ \ 
%\cal{AS}
%\label{Kr11AAa_qPVC}
%\eea

In the discussion above, we mentioned a number of correlations which were neglected in the completed versions ${\cal K}^{r[11]cc}$ and ${\cal K}^{r[11]c}$ of the dynamical kernel.  The major missing correlations are associated with the $\langle 0|\alpha^{\dagger}_{\mu}\alpha_{\mu'}|n\rangle$ amplitudes, the terms of the residual interaction other than $H^{31}$, and the off-diagonal ${\cal K}^{r[12]}$ and ${\cal K}^{r[21]}$ contributions. 
They can be, however, included consistently in the presented framework using a similar technique.

\section{Variational formulation}

There are various ways of solving the EOM (\ref{BSDE}) for the response function. In most of the cases, the desired quantities are the strength functions $S(\omega)$ of Eq. (\ref{SF}), associated with the given external fields $F$ and $F^{\dagger}$ defined by Eq. (\ref{Fext}). Thus, instead of solving the EOM (\ref{BSDE}) for the large matrix of the response function ${\hat{\cal R}}(\omega)$, it can be more convenient to first perform a single contraction of this function with the external field operator, i.e., to introduce the one-body quantity
\bea
\left(\begin{array}{c}\delta{\cal R}^{20}_{\mu\mu'}(\omega) \\ \\ \delta{\cal R}^{02}_{\mu\mu'}(\omega)\end{array}\right) = 
% \frac{1}{2}
 \sum\limits_{\nu\leq\nu'}
\left(\begin{array}{cc} {R}^{[11]}_{\mu\mu'\nu\nu'}(\omega)  &  {R}^{[12]}_{\mu\mu'\nu\nu'}(\omega) 
\\ \\
{R}^{[21]}_{\mu\mu'\nu\nu'}(\omega)  & {R}^{[22]}_{\mu\mu'\nu\nu'}(\omega) \end{array}\right)\nonumber\\
\times\left(\begin{array}{c} F^{20}_{\nu\nu'} \\ \\ F^{02}_{\nu\nu'} \end{array}\right). \ \ \ \ \ \ \ \ \ \ \ \ 
\eea
Applying this operation to Eq. (\ref{BSDE}) yields
\bea
\left(\begin{array}{c}\delta{\cal R}^{20}_{\mu\mu'}(\omega) \\ \\ \delta{\cal R}^{02}_{\mu\mu'}(\omega)\end{array}\right) = 
\left(\begin{array}{c}\delta{\cal R}^{20(0)}_{\mu\mu'}(\omega) \\ \\ \delta{\cal R}^{02(0)}_{\mu\mu'}(\omega)\end{array}\right) 
\ \ \ \ \ \ \ \ \ \ \ \ \ \ \ \ \ \ \ \ \ \ \ \ \ \ \ \ \ \ \ \ \nonumber \\
+ %\frac{1}{4}
\sum\limits_{\nu\nu'\gamma\gamma'}
\large{
\left(\begin{array}{cc} \frac{ {\cal N}^{[11]}_{\mu\mu'\nu\nu'}}{\omega-E_{\mu\mu'}}&
0
\\
0&
\frac{ {\cal N}^{[22]}_{\mu\mu'\nu\nu'}}{\omega+E_{\mu\mu'}}\end{array}\right)
}
\left(\begin{array}{cc} {\cal K}^{[11]}_{\nu\nu'\gamma\gamma'}(\omega) & {\cal K}^{[12]}_{\nu\nu'\gamma\gamma'}(\omega) \\ \\
{\cal K}^{[21]}_{\nu\nu'\gamma\gamma'}(\omega) & {\cal K}^{[22]}_{\nu\nu'\gamma\gamma'}(\omega)\end{array}\right) \nonumber \\
\times\left(\begin{array}{c}\delta{\cal R}^{20}_{\gamma\gamma'}(\omega) \\ \\ \delta{\cal R}^{02}_{\gamma\gamma'}(\omega)\end{array}\right),
\ \ \ \ \ \ \ \ \ \ \ \ 
\nu\leq\nu',\gamma\leq \gamma' , \ \ \ \   \nonumber\\
\label{R2002}
\eea
where
\bea
\left(\begin{array}{c}\delta{\cal R}^{20(0)}_{\mu\mu'}(\omega) \\ \\ \delta{\cal R}^{02(0)}_{\mu\mu'}(\omega)\end{array}\right) = 
\sum\limits_{\nu\leq\nu'}
\large{
\left(\begin{array}{cc}\frac{   {\cal N}^{[11]}_{\mu\mu'\nu\nu'}}{\omega-E_{\mu\mu'}} &
0
\\
0&
\frac{ {\cal N}^{[22]}_{\mu\mu'\nu\nu'}}{\omega+E_{\mu\mu'}}\end{array}\right)
}
\left(\begin{array}{c} F^{20}_{\nu\nu'} \\ \\ F^{02}_{\nu\nu'} \end{array}\right).
%}
\nonumber\\
\eea
The quantities $\delta{\cal R}^{20}$ and $\delta{\cal R}^{02}$ are the components of the density matrix variation in the quasiparticle space. If the ground state is modeled by the HFB approximation,
Eq. (\ref{R2002}) can be recast as follows:
\bea
\delta{\cal R}^{20}_{\mu\mu'}(\omega) = \ \ \ \ \ \ \ \ \ \ \ \ \ \ \ \ \ \ \ \ \ \ \ \ \ \ \ \ \ \ \ \ \ \ \ \ \ \ \ \ \ \ \ \ \ \ \ \ \ \ \ \ \ \ \ \ \ \ \ \ \   \nonumber \\
= \frac{F^{20}_{\mu\mu'} + \sum\limits_{\nu\leq\nu'} \bigl({\cal K}^{[11]}_{\mu\mu'\nu\nu'}(\omega)\delta{\cal R}^{20}_{\nu\nu'}(\omega) + {\cal K}^{[12]}_{\mu\mu'\nu\nu'}(\omega)\delta{\cal R}^{02}_{\nu\nu'}(\omega)\bigr)
}
{\omega - E_{\mu\mu'}} \nonumber \\
\delta{\cal R}^{02}_{\mu\mu'}(\omega) = \ \ \ \ \ \ \ \ \ \ \ \ \ \ \ \ \ \ \ \ \ \ \ \ \ \ \ \ \ \ \ \ \ \ \ \ \ \ \ \ \ \ \ \ \ \ \ \ \ \ \ \ \ \ \ \ \ \ \ \ \   \nonumber \\
= \frac{F^{02}_{\mu\mu'} + \sum\limits_{\nu\leq\nu'} \bigl({\cal K}^{[21]}_{\mu\mu'\nu\nu'}(\omega)\delta{\cal R}^{20}_{\nu\nu'}(\omega) + {\cal K}^{[22]}_{\mu\mu'\nu\nu'}(\omega)\delta{\cal R}^{02}_{\nu\nu'}(\omega)\bigr)
}
{-\omega - E_{\mu\mu'}} . \nonumber \\
\label{FAMgen}
\eea
The latter equation, as well as the more general Eq. (\ref{R2002}) can be interpreted as a generalization of the finite-amplitude QRPA (FAM-QRPA) \cite{Avogadro2011,Avogadro2013,Hinohara2013,Bjelcic2020}. Indeed, the static part $\hat{\cal K}^0$ of the full interaction kernel  ${\hat{\cal K}}(\omega) = {\hat{\cal K}}^0 + {\hat{\cal K}}^r(\omega)$
can be transformed using its relation to the $\cal A$ and $\cal B$ QRPA matrices (\ref{ABN}) \cite{Avogadro2013,Hinohara2013}:
\bea
\delta H^{20}_{\mu\mu'}(\omega) = \sum\limits_{\nu\leq\nu'} \bigl({\cal A}_{\mu\mu',\nu\nu'}\delta{\cal R}^{20}_{\nu\nu'}(\omega) + {\cal B}_{\mu\mu',\nu\nu'}\delta{\cal R}^{02}_{\nu\nu'}(\omega)\bigr) \nonumber\\
- (E_{\mu} + E_{\mu'}) \delta{\cal R}^{20}_{\mu\mu'}(\omega) \nonumber \\
= \sum\limits_{\nu\leq\nu'} \bigl({\cal K}^{0[11]}_{\mu\mu'\nu\nu'}(\omega)\delta{\cal R}^{20}_{\nu\nu'}(\omega) + {\cal K}^{0[12]}_{\mu\mu'\nu\nu'}(\omega)\delta{\cal R}^{02}_{\nu\nu'}(\omega)\bigr)
\nonumber\\
\delta H^{02}_{\mu\mu'}(\omega) = \sum\limits_{\nu\leq\nu'} \bigl({\cal A}^{\ast}_{\mu\mu',\nu\nu'}\delta{\cal R}^{02}_{\nu\nu'}(\omega) + {\cal B}^{\ast}_{\mu\mu',\nu\nu'}\delta{\cal R}^{20}_{\nu\nu'}(\omega)\bigr) \nonumber\\
- (E_{\mu} + E_{\mu'}) \delta{\cal R}^{02}_{\mu\mu'}(\omega) \nonumber\\
= \sum\limits_{\nu\leq\nu'} \bigl({\cal K}^{0[21]}_{\mu\mu'\nu\nu'}(\omega)\delta{\cal R}^{20}_{\nu\nu'}(\omega) + {\cal K}^{0[22]}_{\mu\mu'\nu\nu'}(\omega)\delta{\cal R}^{02}_{\nu\nu'}(\omega)\bigr).
\nonumber\\
\label{dHAB}
\eea
Thus, Eq. (\ref{FAMgen}) can be further transformed as
\bea
\delta{\cal R}^{20}_{\mu\mu'}(\omega) 
= \frac{F^{20}_{\mu\mu'} + \delta H^{20}_{\mu\mu'}(\omega) + \delta{\cal K}^{20}_{\mu\mu'}(\omega)
}
{\omega - E_{\mu\mu'}} \nonumber \\
\delta{\cal R}^{02}_{\mu\mu'}(\omega) 
= \frac{F^{02}_{\mu\mu'} + \delta H^{02}_{\mu\mu'}(\omega) + \delta{\cal K}^{02}_{\mu\mu'}(\omega)
}
{-\omega - E_{\mu\mu'}} , \nonumber \\
\label{FAMgen_fin}
\eea
where
\bea
\delta{\cal K}^{20}_{\mu\mu'}(\omega) = \sum\limits_{\nu\leq\nu'} \bigl({\cal K}^{r[11]}_{\mu\mu'\nu\nu'}(\omega)\delta{\cal R}^{20}_{\nu\nu'}(\omega) \nonumber \\
\label{dK20}
+ {\cal K}^{r[12]}_{\mu\mu'\nu\nu'}(\omega)\delta{\cal R}^{02}_{\nu\nu'}(\omega)\bigr) \\
\delta{\cal K}^{02}_{\mu\mu'}(\omega) = \sum\limits_{\nu\leq\nu'} \bigl({\cal K}^{r[21]}_{\mu\mu'\nu\nu'}(\omega)\delta{\cal R}^{20}_{\nu\nu'}(\omega) \nonumber \\
\label{dK02}
+ {\cal K}^{r[22]}_{\mu\mu'\nu\nu'}(\omega)\delta{\cal R}^{02}_{\nu\nu'}(\omega)\bigr).
\eea  

As it follows from Eqs. (\ref{Gamma11_HFB}, \ref{Dens}, \ref{Vert}), the determination of the qPVC vertices $\Gamma^n$ defining the dynamical kernel ${\cal K}^r$ in the leading approximation requires information about the ${\cal X}^n$
and ${\cal Y}^n$ amplitudes, which can be connected to the superfluid density variations $\delta{\cal R}^{20}$ and $\delta{\cal R}^{02}$ \cite{Hinohara2013,Litvinova2021}. Thereby, the EOM, for instance, in the form of Eq. (\ref{FAMgen_fin}) becomes manifestly non-linear with respect to these amplitudes, as compared to the QRPA, while the non-linearities are generated solely by the dynamical kernel ${\cal K}^r$.  Thus, the finite-amplitude form of the superfluid response EOM creates an attractive opportunity for computing the superfluid density variations and the associated excitation spectra in a single iterative procedure. Such a procedure is, in principle, a powerful extension of the FAM-QRPA beyond the QRPA confined by only the static kernel. The major technical difficulty is the mixing of all the channels (spin, isospin and parities) by the qPVC, while in the QRPA case the channels can be fully decoupled. 

In the implementations with effective interactions, quite a good description of the qPVC vertices of the most important phonons can be achieved on the QRPA  level, i.e., without the dynamical kernel. In this case, a three-step procedure can be performed: %First, 
\begin{enumerate}
%(i) 
\item{The QRPA calculations are run to obtain the ${\cal X}^n$
and ${\cal Y}^n$ amplitudes in various channels;} 
%(ii) 
\item{
The qPVC vertices $\Gamma^n$ are extracted via Eq. (\ref{Gamma11_HFB});% or, in the case of FAM-QRPA in step 1, via Eq. (\ref{Gijcont}) below \cite{Litvinova2021}; 
}
%(iii) 
\item{
The latter vertices enter the dynamical kernel ${\cal K}^r$ in a desired approximation and Eq. (\ref{FAMgen_fin}) is solved to determine the new ${\cal X}^n$
and ${\cal Y}^n$, or $\delta{\cal R}^{20}$ and $\delta{\cal R}^{02}$, amplitudes with the subsequent calculation of the strength distribution in a fixed channel.}
\end{enumerate}
This type of calculation scheme was realized in multiple implementations for spherical nuclei, for example, in Refs. \cite{LitvinovaTselyaev2007,LitvinovaRingTselyaev2008,LitvinovaRingTselyaev2010,LitvinovaSchuck2019}, although not yet via the FAM procedure. Since the effective interactions employed in these implementations contain implicitly the qPVC in a static approximation, the interaction kernel should be corrected to remove the double counting of the qPVC effects. A procedure of subtracting its own value at $\omega = 0$ from the dynamical kernel was formulated in Ref. \cite{Tselyaev2013} and since then is systematically used in beyond-QRPA calculations with effective interactions. Furthermore, in such frameworks, qPVC vertices can be extracted from the FAM-QRPA by taking the variations of the quasiparticle Hamiltonian at the poles of $\delta{\cal R}(\omega)$, as it was shown in Ref. \cite{Litvinova2021} and employed in Ref. \cite{Zhang2022}:
\be
\Gamma^{(ij)n}_{\mu\mu'}  = \frac{1}{\langle n|F^{\dagger}|0\rangle} \oint\limits_{\gamma_n}\delta H^{ij}_{\mu\mu'}(\omega)\frac{d\omega}{2\pi i},
\label{Gijcont}
\ee
where $\gamma_n$ is a contour enclosing the pole $\omega = \omega_n - i\Delta$. 

In the implementations with a bare NN-interaction, more steps may be required. Since in this case the QRPA phonons can be quite unrealistic, they are not likely to produce an adequate approximation to the dynamical kernel. One of the options could be employing an effective interaction on the first two steps, to enter the iteration cycle, and then run step three in a hybrid form with the bare interaction and effective phonons. After that the obtained amplitudes can be recycled until convergence. This and other possibilities will be explored in the future work.

\section{Summary and outlook}
\label{summary}
In this article we presented a response theory for superfluid fermionic systems, which extends beyond the previously existing formulations. Starting from the general many-body Hamiltonian, a consistent derivation of the equation of motion for the superfluid response function was conducted in the HFB basis. 
The resulting EOM is of the Bethe-Salpeter-Dyson type, while its dimension is doubled as compared to the analogous EOM for the response of non-superfluid systems. The interaction kernel splits naturally into the static and dynamical parts, if the underlying bare interaction is of an instantaneous character.
Retaining the static kernel alone leads to the quasiparticle random phase approximation (under the HFB constraint on the ground state), while the dynamical kernel heads beyond the QRPA.

The major emphasis of this work was put on the dynamical kernel, which generates, in principle, a hierarchy of coupled EOMs for increasing-rank fermionic correlation functions. A truncation scheme, which includes completely the superfluid two-fermion correlations, was worked out within a particle number non-conserving framework. Under this truncation, the superfluid response theory is brought to a closed form.
Furthermore, by identifying the correlations, which are most important in the strongly-coupled regime, a leading approximation to the dynamical kernel is obtained in an implementation-ready form. After introducing the notion of superfluid phonons unifying the normal and pairing mutual counterparts, a mapping to the concept of the quasiparticle-phonon coupling  (also referred to as quasiparticle-vibration coupling, although not all the phonons are necessarily of a vibrational character) is achieved. The emergent character of the qPVC is made evident via its explicit link to the bare interaction between the fermions.  Thus, such features as  (i) the link to the bare interaction, (ii) keeping the full HFB character of the pairing correlations in the final expressions and (iii) its systematically improvable character advance the presented approach beyond the existing qPVC models for the nuclear response \cite{Tselyaev2007,Niu2016}. 

The EOM for the response function was further transformed to the equation for the superfluid density matrix variation in a given external field. This allowed us to formulate an extension of the finite-amplitude method FAM, which was up until now confined by the QRPA. As the QRPA is known to provide a very poor picture of the response and associated quantities, in particular, in the nuclear systems, our FAM-qPVC approach, therefore, opens the way to a higher-quality description which is, at the same time, especially efficient because of employing the finite-amplitude technique.

In particular, the response of open-shell deformed nuclei to various external fields can be straightforwardly implemented, thus, broadening the class of nuclear systems described by the qPVC models, which up until now is limited by spherical nuclei. Since the qPVC improves considerably the description of both high-energy and low-energy spectra, it plays an important role in generating accurate rates of various nuclear processes, such as the gamma decay, radiative neutron capture, electron capture, beta decay, and beta-delayed neutron emission. These rates are known to be a key ingredient for modeling astrophysical cataclysmic events of supernova explosions and kilonova, and they are in high demand for both spherical and deformed nuclei. Moreover, accurate nuclear correlation functions in various geometries are necessary for the ongoing searches for the new physics beyond the Standard Model in the nuclear domain, such as the neutrinoless double beta decay and the electric dipole moment.
The numerical implementations of the developed approach and their accommodation for these applications will be addressed by future effort.

%The theory is, thus, adopted for numerical implementations in approximations of growing complexity. 
%GSC 
%ab-initio

%===============================================================================
\section*{Acknowledgement}
Illuminating discussions with Peter Schuck are gratefully acknowledged.
This work was partly supported by the US-NSF CAREER Grant PHY-1654379 and US-NSF Grant PHY-2209376.
Y.Z. acknowledges funding from Guangdong Major Project of Basic and Applied Basic Research under Grant No. 2021B0301030006.
%
%===============================================================================

\appendix
\section{Hamiltonian matrix elements in the quasiparticle space}
\label{AppA}
The matrix elements of the fermionic Hamiltonian (\ref{Hqua}) read, in agreement with Ref. \cite{RingSchuck1980}:
\bea
H^0 &=& \sum\limits_{12}h_{12}\rho_{21} + \frac{1}{2}\sum\limits_{1234}\rho_{31}{\bar v}_{1234}\rho_{42} \nonumber \\
&+& \frac{1}{4}\sum\limits_{1234}\varkappa^{\ast}_{12}{\bar v}_{1234}\varkappa_{34} 
\\
H^{11}_{\mu\nu} &=& \sum\limits_{12}\bigl(U^{\dagger}_{\mu 1}h_{12}U_{2\nu} - V^{\dagger}_{\mu 1}h^T_{12}V_{2\nu} 
 \nonumber \\ &+& U^{\dagger}_{\mu 1}\Delta_{12} V_{2\nu} 
- V^{\dagger}_{\mu 1}\Delta^{\ast}_{12}U_{2\nu}\bigr)
\\
H^{20}_{\mu\nu} &=& \sum\limits_{12}\bigl(U^{\dagger}_{\mu 1}h_{12}V^{\ast}_{2\nu} - V^{\dagger}_{\mu 1}h^T_{12}U^{\ast}_{2\nu} 
\nonumber \\ &+& U^{\dagger}_{\mu 1}\Delta_{12} U^{\ast}_{2\nu}  - V^{\dagger}_{\mu 1}\Delta^{\ast}_{12}V^{\ast}_{2\nu}\bigr)
\\
H^{40}_{\mu\mu'\nu\nu'} &=& \frac{1}{4}\sum\limits_{1234}{\bar v}_{1234} U^{\ast}_{1\mu}U^{\ast}_{2\mu'}V^{\ast}_{4\nu}V^{\ast}_{3\nu'}
\\
H^{31}_{\mu\mu'\nu\nu'} &=& \frac{1}{2}\sum\limits_{1234}{\bar v}_{1234}\bigl(U^{\ast}_{1\mu}V^{\ast}_{4\mu'}V^{\ast}_{3\nu}V_{2\nu'} %\nonumber\\
+ V^{\ast}_{3\mu}U^{\ast}_{2\mu'}U^{\ast}_{1\nu}U_{4\nu'}\bigr) \nonumber\\
\label{H31}
\\
H^{22}_{\mu\mu'\nu\nu'} &=& \sum\limits_{1234}{\bar v}_{1234}\Bigl[\bigl(U^{\ast}_{1\mu}V^{\ast}_{4\mu'}V_{2\nu}U_{3\nu'}
- (\mu \to \mu')\bigr)  \nonumber\\  &-& \bigl(\nu \to \nu'\bigr) 
+ U^{\ast}_{1\mu}U^{\ast}_{2\mu'}U_{3\nu}U_{4\nu'} + V^{\ast}_{3\mu}V^{\ast}_{4\mu'}V_{1\nu}V_{2\nu'}
\Bigr].   \nonumber\\
\eea

\section{Generic commutators}
\label{AppB}
Complementary to the quasiparticle pair operator $A_{\mu\mu'}$ (\ref{Aq}), it is convenient to introduce the operator
\be
C_{\mu\nu} = \alpha^{\dagger}_{\mu}\alpha_{\nu} = C^{\dagger}_{\nu\mu}.
\ee
The first basic commutator is the one between the different quasiparticle pair operators:
\bea
[A_{\mu\mu'},A^{\dagger}_{\nu\nu'}] = -\bigl\{[(\delta_{\mu\nu}C_{\nu'\mu'}) - (\mu\leftrightarrow\mu')] - [(\nu\leftrightarrow\nu')]\bigr\} +\nonumber\\
+ \delta_{\mu\mu'\nu\nu'} = B_{\mu\mu'\nu\nu'} = B^{\dagger}_{\nu\nu'\mu\mu'} = -B_{\mu'\mu\nu\nu'} = -B_{\mu\mu'\nu'\nu}.\nonumber\\
\eea
Furthermore, the commutators entering the interaction kernel read:
\bea
[\alpha^{\dagger}_{\nu}\alpha^{\dagger}_{\nu'}\alpha^{\dagger}_{\gamma}\alpha^{\dagger}_{\gamma'},\alpha_{\mu'}\alpha_{\mu}] \equiv 
[A^{\dagger}_{\nu\nu'}A^{\dagger}_{\gamma\gamma'},A_{\mu\mu'}] \nonumber\\
= -B_{\mu\mu'\nu\nu'}A^{\dagger}_{\gamma\gamma'} - A^{\dagger}_{\nu\nu'}B_{\mu\mu'\gamma\gamma'},
\eea
that is associated with the non-vanishing contribution of $H^{40}$. The term containing $H^{31}$ can be evaluated with
\bea
[\alpha^{\dagger}_{\gamma}\alpha_{\gamma'},A_{\mu\mu'}] &=& [C_{\gamma\gamma'},A_{\mu\mu'}] \nonumber\\
= \delta_{\gamma\mu}\alpha_{\gamma'}\alpha_{\mu'} - \delta_{\gamma\mu'}\alpha_{\gamma'}\alpha_{\mu} 
&\equiv& \delta_{\gamma\mu}A_{\mu'\gamma'} - \delta_{\gamma\mu'}A_{\mu\gamma'} 
= D_{\gamma\gamma'\mu\mu'}
\nonumber\\
\eea
and, further,
\bea
[A^{\dagger}_{\nu\nu'}\alpha^{\dagger}_{\gamma}\alpha_{\gamma'},A_{\mu\mu'}] \equiv [A^{\dagger}_{\nu\nu'}C_{\gamma\gamma'},A_{\mu\mu'}]
\nonumber\\
= A^{\dagger}_{\nu\nu'}D_{\gamma\gamma'\mu\mu'} - B_{\mu\mu'\nu\nu'}C_{\gamma\gamma'},
\eea
\bea
[\alpha^{\dagger}_{\gamma'}\alpha_{\gamma}A_{\nu\nu'},A_{\mu\mu'}] \equiv [C_{\gamma'\gamma}A_{\nu\nu'},A_{\mu\mu'}] 
\nonumber\\
= D_{\gamma'\gamma\mu\mu'}A_{\nu\nu'},
\eea
\bea
[A^{\dagger}_{\nu\nu'}\alpha^{\dagger}_{\gamma}\alpha_{\gamma'},A^{\dagger}_{\mu\mu'}] \equiv [A^{\dagger}_{\nu\nu'}C_{\gamma\gamma'},A^{\dagger}_{\mu\mu'}] 
\nonumber\\
=  -A^{\dagger}_{\nu\nu'}D^{\dagger}_{\gamma'\gamma\mu\mu'}.
\eea
\bea
[\alpha^{\dagger}_{\gamma'}\alpha_{\gamma}A_{\nu\nu'},A^{\dagger}_{\mu\mu'}] \equiv [C_{\gamma'\gamma}A_{\nu\nu'},A^{\dagger}_{\mu\mu'}] 
\nonumber\\
= C_{\gamma'\gamma}B^{\dagger}_{\mu\mu'\nu\nu'} -D^{\dagger}_{\gamma\gamma'\mu\mu'}A_{\nu\nu'}.
\eea
Finally, the commutators associated with $H^{22}$ read:
\bea
[A^{\dagger}_{\gamma\gamma'}A_{\nu\nu'},A_{\mu\mu'}] = -B_{\mu\mu'\gamma\gamma'}A_{\nu\nu'},
\nonumber
\\
\left[A^{\dagger}_{\gamma\gamma'}A_{\nu\nu'},A^{\dagger}_{\mu\mu'}\right] = A^{\dagger}_{\gamma\gamma'}B_{\nu\nu'\mu\mu'}.
\eea

Thus, the first commutator of Eqs. (\ref{T0}) is given by:
\bea
[V,A_{\mu\mu'}] = -\sum\limits_{\nu\nu'\gamma\gamma'} H^{40}_{\nu\nu'\gamma\gamma'} (B_{\mu\mu'\nu\nu'}A^{\dagger}_{\gamma\gamma'} + 
A^{\dagger}_{\nu\nu'}B_{\mu\mu'\gamma\gamma'}) \nonumber\\
+\sum\limits_{\nu\nu'\gamma\gamma'} \bigl[H^{31}_{\nu\nu'\gamma\gamma'}(A^{\dagger}_{\nu\nu'}D_{\gamma\gamma'\mu\mu'} - B_{\mu\mu'\nu\nu'}C_{\gamma\gamma'}) \nonumber\\
+ H^{31\ast}_{\nu\nu'\gamma\gamma'}D_{\gamma'\gamma\mu\mu'}A_{\nu\nu'}\bigr] \nonumber\\
- \frac{1}{4}\sum\limits_{\nu\nu'\gamma\gamma'} \bigl(H^{22}_{\nu\nu'\gamma\gamma'}B_{\mu\mu'\nu\nu'}A_{\gamma\gamma'} + H^{22\ast}_{\nu\nu'\gamma\gamma'}B_{\mu\mu'\gamma\gamma'}A_{\nu\nu'}\bigr),\nonumber\\
\label{VAcomm}
\eea
while the second one can be deduced from it by Hermitian conjugation:
\be
[V,A^{\dagger}_{\mu\mu'}] = -[V,A_{\mu\mu'}]^{\dagger},
\label{VAdcomm}
\ee
since $V = V^{\dagger}$, i.e., is Hermitian.

\section{Contributions to the dynamical kernel}
\label{AppC}
Applying the Fourier transformation to the terms of type (a) 
of Eq. (\ref{Kr11AA}) leads to 
\bea
{\cal K}^{r[11]AAa}_{\mu\mu'\nu\nu'}(\omega) &=& \frac{1}{4}
%\Bigl\{\Bigl[
\sum\limits_{\gamma\delta nm}\Bigl[\frac{\theta^n_{\mu\gamma}{\cal X}^{m}_{\mu'\gamma}{\cal X}^{m\ast}_{\nu'\delta}\theta^{n\ast}_{\nu\delta}}{\omega - \omega_{nm} + i\delta} \nonumber\\ &-& 
\frac{\theta^{n\ast}_{\gamma\mu}{\cal Y}^{m\ast}_{\mu'\gamma}{\cal Y}^{m}_{\nu'\delta}\theta^{n}_{\delta\nu}}{\omega + \omega_{nm} - i\delta}\Bigr]
%\Bigr]
- \cal{AS},
\label{Kr11AAa}
\eea
where the antisymmetrization $\cal{AS}$ is implied with respect to $\mu\leftrightarrow\mu'$ and $\nu\leftrightarrow\nu'$ and the vertex functions are introduced according to:
\be
\theta^n_{\mu\gamma} = 2\sum\limits_{\rho\rho'}(H^{31\ast}_{\rho\rho'\gamma\mu}{\cal X}^n_{\rho\rho'} + H^{31}_{\rho\rho'\mu\gamma}{\cal Y}^n_{\rho\rho'}).
%\nonumber\\
%= \sum\limits_{1234}\sum\limits_{\rho\rho'}\bigl[{\bar v}^{\ast}_{1234}(U_{1\rho}V_{4\rho'}V_{3\gamma}V^{\ast}_{2\mu}
%+ V_{3\rho}U_{2\rho'}U_{1\gamma}U^{\ast}_{4\mu}){\cal X}^n_{\rho\rho'} \nonumber\\
%+ {\bar v}_{1234}(U^{\ast}_{1\rho}V^{\ast}_{4\rho'}V^{\ast}_{3\mu}V_{2\gamma}
%+ V^{\ast}_{3\rho}U^{\ast}_{2\rho'}U^{\ast}_{1\mu}U_{4\gamma}){\cal Y}^n_{\rho\rho'}
%\bigr]
%\nonumber\\
%= \sum\limits_{12}(U^{\dagger}_{\mu 1}g^n_{12}U_{2\gamma} - V^{\dagger}_{\mu 1}g^{nT}_{12}V_{2\gamma}).
%\nonumber\\
%\label{theta_n}
\label{vert1}
\ee

Another type of contributions is associated with the permutations of the quasiparticle operators $\alpha_{\mu}$ and $\alpha^{\dagger}_{\mu}$ belonging to different 
two-quasiparticle pairs. Proceeding similarly with the antisymmetrizations $\eta \leftrightarrow \delta, \eta \leftrightarrow \nu', \rho \leftrightarrow \gamma$, and $\rho \leftrightarrow \mu'$ in Eqs. (\ref{Tr11AA2} - \ref{Tr11AA4}) between the same kinds of the two-quasiparticle operators at the same times leads to the following contribution:
\bea
{\cal K}^{r[11]AAa;x}_{\mu\mu'\nu\nu'}(\omega) = \frac{1}{4}
%\Bigl\{\Bigl[
\sum\limits_{\gamma\delta nm}\Bigl[\frac{{\cal X}^{m}_{\mu'\gamma}{\cal X}^{m\ast}_{\nu'\delta}}{\omega - \omega_{nm} + i\delta} \ \ \ \ \ \ \ \ \ \ 
\nonumber\\
\times\bigl(\theta^n_{\mu\gamma}{\bar\xi}^{n\ast}_{\nu\delta} + {\bar\xi}^n_{\mu\gamma}{\theta}^{n\ast}_{\nu\delta} + {\bar\xi}^n_{\mu\gamma}{\bar\xi}^{n\ast}_{\nu\delta} \bigr)
%+ \theta^n_{\mu\gamma}{\xi}^{n\ast}_{\nu\delta} + {\xi}^n_{\mu\gamma}{\theta}^{n\ast}_{\nu\delta} + {\xi}^n_{\mu\gamma}{\xi}^{n\ast}_{\nu\delta}\bigr)
\nonumber\\ - 
\frac{{\cal Y}^{m\ast}_{\mu'\gamma}{\cal Y}^{m}_{\nu'\delta}}{\omega + \omega_{nm} - i\delta} \ \ \ \ \ \ \ \ \ \ 
\nonumber\\
\times\bigl(
%\theta^{n\ast}_{\gamma\mu}{\bar\xi}^{n}_{\delta\nu} + {\bar\xi}^{n\ast}_{\gamma\mu}{\theta}^{n}_{\delta\nu} + {\bar\xi}^{n\ast}_{\gamma\mu}{\bar\xi}^{n}_{\delta\nu} + 
%
\theta^{n\ast}_{\gamma\mu}{\xi}^{n}_{\delta\nu} + {\xi}^{n\ast}_{\gamma\mu}{\theta}^{n}_{\delta\nu} + {\xi}^{n\ast}_{\gamma\mu}{\xi}^{n}_{\delta\nu}\bigr)
\Bigr] \nonumber\\
%\Bigr]
- \cal{AS}, \ \ \ \ \ \ \ \ \ \ 
\label{Kr11AAax}
\eea
where ${\xi}^{n}$ and ${\bar\xi}^{n}$ stand for the combinations:
\bea
\xi^n_{\mu\gamma} = 2\sum\limits_{\rho\rho'}(H^{31}_{\mu\rho\rho'\gamma} - H^{31}_{\rho\mu\rho'\gamma}){\cal Y}^n_{\rho\rho'}\nonumber\\
{\bar\xi}^n_{\delta\nu} = 2\sum\limits_{\rho\rho'}(H^{31\ast}_{\nu\rho\rho'\delta} - H^{31\ast}_{\rho\nu\rho'\delta}){\cal X}^n_{\rho\rho'}.
\label{vert23}
\eea

Furthermore, the contributions of type (a) from the $C$ operators without breaking two-quasiparticle pairs read:
\bea
{\cal K}^{r[11]CCa}_{\mu\mu'\nu\nu'}(\omega) &=& \frac{1}{4}
%\Bigl\{\Bigl[
\sum\limits_{\gamma\delta nm}\Bigl[\frac{\xi^n_{\mu\gamma}{\cal X}^{m}_{\mu'\gamma}{\cal X}^{m\ast}_{\nu'\delta}\xi^{n\ast}_{\nu\delta}}{\omega - \omega_{nm} + i\delta} \nonumber\\ &-& 
\frac{{\bar\xi}^{n\ast}_{\gamma\mu}{\cal Y}^{m\ast}_{\mu'\gamma}{\cal Y}^{m}_{\nu'\delta}{\bar\xi}^{n}_{\delta\nu}}{\omega + \omega_{nm} - i\delta}\Bigr]
%\Bigr]
- \cal{AS},
\label{Kr11CCa}
\eea
while the analogous contributions from the products of the $C$-operator and $A$-operator terms are the following:
\bea
{\cal K}^{r[11]ACa}_{\mu\mu'\nu\nu'}(\omega) = \frac{1}{4}
\sum\limits_{\gamma\delta nm}\Bigl[\frac{{\cal X}^{m}_{\mu'\gamma}{\cal X}^{m\ast}_{\nu'\delta}}{\omega - \omega_{nm} + i\delta} 
\bigl(\theta^n_{\mu\gamma}{\xi}^{n\ast}_{\nu\delta} + {\xi}^n_{\mu\gamma}{\theta}^{n\ast}_{\nu\delta} \bigr)
\nonumber\\ - 
\frac{{\cal Y}^{m\ast}_{\mu'\gamma}{\cal Y}^{m}_{\nu'\delta}}{\omega + \omega_{nm} - i\delta}
\bigl(\theta^{n\ast}_{\gamma\mu}{\bar\xi}^{n}_{\delta\nu} + {\bar\xi}^{n\ast}_{\gamma\mu}{\theta}^{n}_{\delta\nu}\bigr)
\Bigr] \nonumber\\
%\Bigr]
- \cal{AS}. \ \ \ \ \ \ \ \ \ \ 
\label{Kr11ACa}
\eea
Finally, the cross-pair counterparts of $ {\cal K}^{r[11]ACa}_{\mu\mu'\nu\nu'}(\omega)$ read:
\bea
{\cal K}^{r[11]ACa;x}_{\mu\mu'\nu\nu'}(\omega) = \frac{1}{4}
\sum\limits_{\gamma\delta nm}\Bigl[\frac{{\cal X}^{m}_{\mu'\gamma}{\cal X}^{m\ast}_{\nu'\delta}}{\omega - \omega_{nm} + i\delta} 
\bigl(\xi^n_{\mu\gamma}{\bar\xi}^{n\ast}_{\nu\delta} + {\bar\xi}^n_{\mu\gamma}{\xi}^{n\ast}_{\nu\delta} \bigr)
\nonumber\\ - 
\frac{{\cal Y}^{m\ast}_{\mu'\gamma}{\cal Y}^{m}_{\nu'\delta}}{\omega + \omega_{nm} - i\delta}
\bigl({\bar\xi}^{n\ast}_{\gamma\mu}{\xi}^{n}_{\delta\nu} + {\xi}^{n\ast}_{\gamma\mu}{\bar\xi}^{n}_{\delta\nu}\bigr)
\Bigr] \nonumber\\
%\Bigr]
- \cal{AS}, \ \ \ \ \ \ \ \ \ \ 
\label{Kr11ACax}
\eea
while the pure $C$-operator terms do not admit further cross-pair antisymmetrizations.

Here we notice that all the terms listed in Eqs. (\ref{Kr11AAa}, \ref{Kr11AAax}, \ref{Kr11CCa}, \ref{Kr11ACa}, \ref{Kr11ACax}) have the same poles and partly the residues, differing only by the multipliers associated with the "partial" vertex functions $\theta^n, \xi^n$ and ${\bar\xi}^n$. Summing them up altogether leads to Eq. (\ref{Kr11cc}).

%===============================================================================

\bibliography{BibliographyNov2021}
\end{document}